\documentclass[10pt,letterpaper]{article}
\usepackage[T1]{fontenc}
\usepackage[utf8]{inputenc}
\usepackage{times}
\usepackage[left=1in,right=1in,top=0.9in,bottom=0.9in]{geometry}
\usepackage{amsmath,amssymb,amsthm,mathtools}
\usepackage{microtype,booktabs,array,enumitem}
\usepackage{xcolor}
\usepackage[colorlinks=true,linkcolor=blue!40!black,citecolor=blue!40!black,urlcolor=blue!40!black]{hyperref}
\usepackage{appendix}
\hypersetup{pdftitle={A new quantum algorithm framework for integer order Renyi entropy estimation},pdfauthor={},pdfsubject={Integer order Renyi entropy estimation under purified quantum query-access},pdfkeywords={Quantum query complexity, Renyi entropy, integer orders, collision states}}
\newtheoremstyle{paper}{6pt}{6pt}{\normalfont}{}{\bfseries}{.}{.5em}{}
\theoremstyle{paper}
\newtheorem{theorem}{Theorem}
\newtheorem{lemma}{Lemma}
\newtheorem{proposition}{Proposition}
\newtheorem{corollary}{Corollary}
\newtheorem{definition}{Definition}
\newtheorem{problem}{Problem}

\newcounter{algorithm}
\newenvironment{algorithmblock}[1]{%
 \par\medskip\noindent\begin{minipage}{\linewidth}\refstepcounter{algorithm}%
 \hrule\vspace{5pt}\textbf{Algorithm \thealgorithm: #1}\par\smallskip\small
}{\vspace{5pt}\hrule\end{minipage}\par\medskip}
\newcommand{\Oh}{\widetilde O}
\newcommand{\eps}{\epsilon}
\newcommand{\ket}[1]{\lvert #1\rangle}
\newcommand{\bra}[1]{\langle #1\rvert}

\newcommand{\norm}[1]{\lVert #1\rVert}
\newcommand{\E}{\mathbb E}
\newcommand{\Var}{\operatorname{Var}}
\newcommand{\Act}{\mathcal A}
\newcommand{\Src}{\mathcal B}
\newcommand{\ind}{\mathbf 1}
\newcommand{\Renyi}{R\'enyi}
\DeclareMathOperator{\spanop}{span}
\renewcommand{\thesection}{\Roman{section}}

\setlist[enumerate]{leftmargin=*,itemsep=3pt,topsep=4pt}
\makeatletter
\renewcommand{\@seccntformat}[1]{\csname the#1\endcsname.\enspace}
\renewcommand\section{\@startsection{section}{1}{\z@}{2.3ex plus 1ex minus .2ex}{1.0ex plus .2ex}{\normalfont\normalsize\centering\scshape}}
\renewcommand\subsection{\@startsection{subsection}{2}{\z@}{1.8ex plus 1ex minus .2ex}{.7ex plus .2ex}{\normalfont\normalsize\itshape}}
\renewcommand{\@maketitle}{\newpage\null\vskip 1em\begin{center}{\fontsize{21}{25}\selectfont\@title\par}\vskip 2em{\large\@author\par}\end{center}\par\vskip .5em}
\makeatother

\begin{document}
\title{A New Quantum Algorithm Framework for \\ Integer Order Entropy Estimation}
\author{Zongxing Xiong}
\date{}

\maketitle
\begin{abstract}
Entropy estimation for unknown distributions is a basic problem in statistical property estimation. In this paper, we propose a quantum algorithm framework for estimating R\'enyi entropy with fixed integer order $k\ge2$. Given purified quantum query-access to a distribution on $n$ symbols, our algorithm estimates its order-$k$ R\'enyi entropy to within additive error $\eps$ using
$$
 \Oh\left(\frac{n^{u_k}}{\eps^{1-u_k}} + \frac{n^{(k-1)/(2k)}}\eps\right)
$$
queries, where $u_k=\max_{2\le j\le k} \frac{(3k-j)2^{j-2}-k}{k(2^j-1)}$.
The main ingredient is a recursive construction of collision states in which intermediate states are retained and reused. We bound both the preparation and reflection costs, and choose the number of retained states at each order so that all recursive steps fit within a common query budget. We combine this construction with a sequence of decreasing probability thresholds, estimating the contribution of the labels found along the way separately from the remaining power sum. A batch estimator gives the required bound in the high-precision regime, and an initial search removes the need for prior knowledge of the power sum. 
\end{abstract}

\medskip
\section{Introduction}
Estimating properties of unknown probability distributions is a fundamental task in statistics and theoretical computer science. Entropy estimation is a particularly important example: it asks how much information about the uncertainty of a distribution can be obtained without reconstructing all of its probabilities. For a distribution $p=(p_i)_{i=1}^n$ on $[n]$ and an integer $k\ge2$, the R\'enyi entropy~\cite{Renyi} and the associated power sum are
\begin{equation}\label{eq:entropy}
 H_k(p):=\frac{1}{1-k}\log P_k(p),
 \qquad P_k(p):=\sum_{i=1}^n p_i^k.
\end{equation}
For integer orders, the power sum admits a useful probabilistic interpretation:
\[
 P_k(p)=\Pr\{X_1=\cdots=X_k\},
 \qquad X_1,\ldots,X_k\overset{\mathrm{iid}}\sim p.
\]
Thus it is the probability that $k$ independent samples have the same label. Such an observation underlies classical collision estimators~\cite{AOST} and quantum algorithms based on distinctness testing~\cite{LiWu}. It also suggests preparing coherent quantum states whose label probabilities are proportional to higher powers of $p_i$ via coherent access to a sampling procedure~\cite{LiWu,GL}.

This paper studies the following estimation problem.

\begin{problem}[Integer order \Renyi{} entropy estimation]\label{prob:entropy}
Given purified quantum query-access to an unknown probability distribution $p$ on $[n]$, a fixed integer $k\ge2$, an additive accuracy $\eps>0$, and a maximum failure probability $\delta\in(0,1/3)$, output an estimate $\widetilde H_k(p)$ satisfying
\[
 \Pr\{|\widetilde H_k(p)-H_k(p)|\le\eps\}\ge1-\delta.
\]
The resource measure is the number of queries to the preparation oracle.
\end{problem}

Two issues arise when using collision states to estimate entropy. First, additive accuracy in $H_k(p)$ requires relative accuracy in $P_k(p)$, whose value may be as small as $n^{1-k}$. Second, a collision state of order $s+1$ is prepared using states of order $s$, so repeated use of the construction must account for the cost of these intermediate states. We address both issues by combining a recursive preparation procedure with quantum mean estimation and a sequence of decreasing probability thresholds.

\subsection{Main results}
We work in the purified quantum query-access model of Definition~\ref{def:access}. For each fixed integer $k\ge2$, define
\begin{equation}\label{eq:exponents}
 u_{k,j}:=\frac{(3k-j)2^{j-2}-k}{k(2^j-1)},
 \qquad u_k:=\max_{2\le j\le k}u_{k,j},
 \qquad b_k:=\frac{k-1}{2k}.
\end{equation}
Our main result is the following theorem.

\begin{theorem}[Main theorem]\label{thm:main}
For any fixed integer $k\ge2$, there is a constant $c_k>0$ and a quantum algorithm such that, for any $n\ge2$, $\eps\in(0,c_k]$, $\delta\in(0,1/3)$, and probability distribution $p$ on $[n]$, the algorithm estimates $H_k(p)$ to within additive error $\eps$ with success probability at least $1-\delta$, using
\begin{equation}\label{eq:main}
 \Oh_k\!\left(\frac{n^{u_k}}{\eps^{1-u_k}}
                         +\frac{n^{b_k}}\eps\right)
\end{equation}
queries to $U_p$ and $U_p^\dagger$ in Definition~\ref{def:access} (controlled access to $U_p$ is not required).
\end{theorem}

 We first construct an estimator with success probability at least $2/3$; independent repetition followed by taking the median gives the stated dependence on $\delta$. The exponents of the query complexity $Q_k(n,\eps)$ for success probability at least $2/3$ are summarized in Table~\ref{tab:orders}. The two terms in Eq.~\eqref{eq:main} are equal when $\eps=n^{-\theta_k}$, where
\begin{equation}\label{eq:theta}
 \theta_k:=1-\frac{b_k}{u_k}.
\end{equation}
Consequently, Theorem~\ref{thm:main} gives
\begin{equation}\label{eq:regimes}
 Q_k(n,\eps)\le
 \begin{cases}
 \Oh_k(n^{u_k}/\eps^{1-u_k}),&n^{-\theta_k}\le\eps\le c_k,\\
 \Oh_k(n^{b_k}/\eps),&0<\eps\le\min\{c_k,n^{-\theta_k}\}.
 \end{cases}
\end{equation}

\medskip
\begin{table}[ht]
\centering
\begin{tabular}{ccccc}
\toprule
Order $k$&$u_k$&$1-u_k$&$b_k$&$\theta_k$\\
\midrule
2&$1/3$&$2/3$&$1/4$&$1/4$\\
3&$4/9$&$5/9$&$1/3$&$1/4$\\
4&$1/2$&$1/2$&$3/8$&$1/4$\\
5&$19/35$&$16/35$&$2/5$&$5/19$\\
6&$4/7$&$3/7$&$5/12$&$13/48$\\
10&$22/35$&$13/35$&$9/20$&$25/88$\\
11&$7/11$&$4/11$&$5/11$&$2/7$\\
12&$29/45$&$16/45$&$11/24$&$67/232$\\
26&$9/13$&$4/13$&$25/52$&$11/36$\\
\bottomrule
\end{tabular}
\caption{Exponents in quantum query upper bounds $Q_k(n,\eps) = \Oh_k(n^{u_k}/\eps^{1-u_k}+n^{b_k}/\eps)$.}\label{tab:orders}
\end{table}

\subsection{Dependence on the integer order}
We describe how the query exponent depends on the integer order. By subtracting adjacent terms in Eq.~\eqref{eq:exponents}, we obtain
\[
 u_{k,j+1}\ge u_{k,j}\quad\Longleftrightarrow\quad
 k\ge2^{j+1}-j-2.
\]
A maximizing level $j$ therefore satisfies
\begin{equation}\label{eq:activelevel}
 2^j-j-1\le k\le2^{j+1}-j-2,
 \qquad 2\le j\le k.
\end{equation}
Adjacent levels give the same value at a shared endpoint. The first ranges are displayed in Table~\ref{tab:levels}. Eq.~\eqref{eq:activelevel} implies $j=\log_2k+O(1)$. Using
\[
 u_{k,j}=\frac34-
 \frac{j2^{j-2}+k/4}{k(2^j-1)}
\]
therefore gives
\begin{equation}\label{eq:asymptotic}
 u_k=\frac34-\frac{\log_2k}{4k}+O(1/k).
\end{equation}
This maximizing level identifies the term that determines the common query budget. Our algorithm still constructs collision prefixes successively up to the requested order $k$.

\medskip
\begin{table}[ht]
\centering
\begin{tabular}{ccc}
\toprule
Level $j$&Integer orders $k$&$u_k$\\
\midrule
2&$2\le k\le4$&$(2k-2)/(3k)$\\
3&$4\le k\le11$&$(5k-6)/(7k)$\\
4&$11\le k\le26$&$(11k-16)/(15k)$\\
5&$26\le k\le57$&$(23k-40)/(31k)$\\
6&$57\le k\le120$&$(47k-96)/(63k)$\\
\bottomrule
\end{tabular}
\caption{Levels controlling the recursive budget. Shared endpoints have two maximizing levels.}\label{tab:levels}
\end{table}

\subsection{Techniques}
We describe the main ideas before giving the constructions and proofs. The framework has three components: a recursive procedure for preparing collision states, a threshold-based decomposition of the power sum, and a preliminary estimate of its scale.

\noindent\textit{1) Recursive preparation of collision states:}
For a known set $\Act\subseteq[n]$, a collision state of order $s$ has label probabilities proportional to $p_i^s$ on $\Act$. To extend it, we retain several copies, choose one uniformly, and compare its label with that of a fresh purified sample. Conditional on equality, the selected registers form a collision state of order $s+1$. The routing operation moves the complete purification registers. Consequently, the unused copies are restored and factor from the output state.

To analyze preparation and reflection together, we use the transducer framework of Belovs, Jeffery, and Yolcu~\cite{BJY} and the amplification and estimation tools of Cornelissen, Apers, and Gribling~\cite{CAG}. We compose the operations of a stage before converting them to a query circuit, so the intermediate states are reused exactly within the transducer construction. Let $M_s=\sum_{i\in\Act}p_i^s$, let $D_s$ bound the preparation and reflection costs at order $s$, and set $E_s^2=M_sD_s^2$. Retaining approximately $B/D_s$ copies gives the recurrence
\begin{equation}\label{eq:intro-rec}
 E_{s+1}^2\le O_k\!\left(
 \varphi E_s^2+\frac{\sqrt{M_s}\,E_s}{B}+\varphi^s\right),
\end{equation}
where $p_i\le\varphi$ on $\Act$. The two recursive terms account for extending a state and for using the retained copies. By expanding the recurrence and bounding each resulting term, we obtain a budget that is sufficient at every order up to $k$. The largest of these requirements determines the exponent $u_k$ in Eq.~\eqref{eq:exponents}.

\noindent\textit{2) Estimation over probability intervals:}
Suppose that $R$ is a known upper bound within a constant factor of $P_k(p)$. Starting from $\varphi_0=R^{1/k}$, we decrease the threshold geometrically, with $\varphi_j=\varphi_0/2^j$. At each stage, samples from the order-$k$ collision state reveal all labels whose probabilities exceed the next threshold, with high probability. We record the observed labels and remove them from the active set. Once the threshold reaches a constant multiple of $(\eta R)^{1/k}$, we estimate the remaining power sum to additive error $\Delta=\Theta_k(\eta R)$, where $\eta$ is the desired relative accuracy.

At most $\Oh_k(1/\eta)$ labels are recorded. We estimate their total contribution by conditioning on this known set and applying a batch estimator. The same estimator is applied directly to the full distribution in the high-precision regime. These two uses of batch estimation account for the second term in Eq.~\eqref{eq:main}. The organization by probability thresholds is related to the multilevel approach of Chen et al.~\cite{Multi}; here the states used at each threshold are constructed from integer-order collisions.

\noindent\textit{3) Removing prior knowledge and the order-two case:}
We obtain $R$ by starting from the upper bound $1$ and halving the candidate scale whenever an additive estimate certifies that the next candidate is still valid. Each test uses a collision-table estimator based on the search algorithm of Peng, Hou, and Dong~\cite{PHD}. The total cost of this preliminary search is absorbed in Eq.~\eqref{eq:main}. For $k=2$, the first mean-estimation step of the collision-state construction suffices for both the preliminary search and the final estimate, giving a particularly simple implementation.

\subsection{Related work}
Collision statistics are a standard tool for estimating integer power sums. Acharya, Orlitsky, Suresh, and Tyagi~\cite{AOST} used this structure in classical R\'enyi entropy estimation. In the quantum setting, Gily\'en and Li~\cite{GL} developed a framework for distributional property testing under purified quantum query-access. Li and Wu~\cite{LiWu} developed entropy estimators for a broad range of orders and reduced the integer-order problem to distinctness testing. Their work provides the starting point for exploiting collision structure in quantum entropy estimation. Our algorithm uses the collision interpretation at integer orders to construct the states needed for estimation recursively.

Quantum amplitude estimation~\cite{BHMT} and variance-dependent mean estimation~\cite{Montanaro} supply the basic estimation procedures. The probability-oracle formulation and multidimensional estimation results of van Apeldoorn~\cite{vA} give further context for coherent access to distributions. Chen, Wang, and Zhang~\cite{CWZ} discuss purified access for a broad collection of distributional and quantum-state properties. We use this access model with arbitrary purification states and count calls to the preparation oracle and its inverse.

Our collision-table search uses learning-graph techniques~\cite{Belovs} in the form developed by Peng, Hou, and Dong~\cite{PHD}. The recursive state construction uses the transducer composition framework~\cite{BJY,CAG}, which is closely related to the query-complexity characterization of state conversion~\cite{LMRSS}. We give the preparation, reflection, and normalization analysis needed to apply these tools to the entropy estimator.

\subsection{Organization}
The rest of the paper is organized as follows. Section~\ref{sec:model} introduces the notation and access model. Section~\ref{sec:batch} gives the batch estimator, and Section~\ref{sec:scale} obtains a coarse power-sum estimate without prior knowledge. In Section~\ref{sec:prefix}, we construct the recursive collision-state primitive. Section~\ref{sec:budgets} analyzes its query complexity. Section~\ref{sec:algorithm} combines these ingredients into the entropy estimator and proves Theorem~\ref{thm:main}.The collision-table estimator used in the binary search is proved in \ref{app:sparse} and a simpler sufficient algorithm for order two that uses only the first recursive step is given in \ref{app:two}.

\medskip
\section{Preliminaries}\label{sec:model}
\subsection{Notation}
Throughout the paper, $k$ is a fixed integer. The notation $\Oh_k$ suppresses polylogarithmic factors in $n$, $1/\eps$, and $1/\delta$, and allows its implicit constants to depend on $k$. All query bounds are worst-case bounds. For an integer $n\ge2$, we write $[n]=\{1,2,\ldots,n\}$. Unless stated otherwise, logarithms are natural and vector norms are Euclidean. The Hilbert space of a quantum register $X$ is denoted by $\mathcal H_X$, and $\ket0$ denotes its all-zero state. For a fixed integer $k\ge2$, $H_k(p)$ and $P_k(p)$ denote the entropy and power sum in Eq.~\eqref{eq:entropy}. The scalar $R$ always denotes a known upper bound for the full power sum and $P$ abbreviates the unknown value $P_k(p)$ when convenient. Estimates are denoted by a tilde. For a fixed set $\Act$, we write $M_s=\sum_{i\in\Act}p_i^s$ for the restricted power sum. Probability thresholds are denoted by $\varphi$ or $\varphi_j$ and are imposed directly on the probabilities $p_i$.

\subsection{Purified quantum query-access}
We use the purified-access model of~\cite{GL}, also formulated as a quantum probability oracle in~\cite{vA}. The following definition specifies the operations charged in our query bounds.

\begin{definition}[Purified quantum query-access for discrete distributions]\label{def:access}
An unknown discrete distribution $p=(p_i)_{i=1}^n$ has purified quantum query-access if we can query a unitary $U_p$ and its inverse such that
\begin{equation}\label{eq:oracle}
 U_p\ket0_B\ket0_C=\ket{\Psi_p}
   =\sum_{i=1}^n\sqrt{p_i}\ket{\phi_i}_B\ket i_C,
\end{equation}
where the states $\ket{\phi_i}$ are arbitrary normalized states. One application of $U_p$ or $U_p^\dagger$ is one query. The action of $U_p$ outside $\ket0_B\ket0_C$ is unrestricted.
\end{definition}
The label register makes the states $\ket{u_i}:=\ket{\phi_i}_B\ket i_C$ orthonormal. All permutations in our algorithms move complete copies of registers $BC$, so the unknown purification states are retained together with their labels.

Controlled versions of the preparation oracle are often included in the access model~\cite{CWZ}. The algorithms below require only $U_p$ and $U_p^\dagger$. Indeed, a controlled reflection about $\Psi_p$ is implemented by applying $U_p^\dagger$, a controlled phase on the known state $\ket0_B\ket0_C$, and then $U_p$. The two oracle calls cancel when the control is inactive. We use the same construction for controlled table lookups.

\subsection{Power sums and entropy estimation}
Let $P=P_k(p)$. The following estimates will be used repeatedly:
\begin{align}
 n^{1-k}\le P\le1,\qquad p_{\max}&\le P^{1/k},\label{eq:basic}\\
 P_s(p)&\le P^{s/k}&& (s\ge k),\label{eq:highmoment}\\
 P_s(p)&\le n^{1-s/k}P^{s/k}&& (0<s\le k),\label{eq:dimensionmoment}\\
 P_s(p)&\le P^{(s-1)/(k-1)}&& (1\le s\le k).\label{eq:interpolation}
\end{align}
The first three inequalities follow from standard comparisons between finite-dimensional norms. To prove Eq.~\eqref{eq:interpolation}, take $X\sim p$ and apply the moment inequality to the nonnegative random variable $p_X$, using $P_s(p)=\E p_X^{s-1}$ and $P_k(p)=\E p_X^{k-1}$.

Suppose that an estimate $\widetilde P$ satisfies $|\widetilde P-P|\le\eta P$ for $\eta\le1/2$. Then
\begin{equation}\label{eq:log}
 \left|\frac{\log\widetilde P}{1-k}-H_k(p)\right|
 \le\frac{2\eta}{k-1}.
\end{equation}
Consequently, relative accuracy $\eta=\Theta_k(\eps)$ in the power sum suffices for additive accuracy $\eps$ in the entropy, as in~\cite{AOST,LiWu}. If only an upper bound $R\le C_k P$ is available, it is enough to request additive accuracy $\Delta=c_k\eps R$ for a sufficiently small constant $c_k$. Before taking the logarithm, the estimate is projected onto $[n^{1-k},1]$. This projection cannot increase its distance from $P$.

\medskip
\section{Estimating power sums with batches of samples}\label{sec:batch}
We first estimate $P_k(p)$ when a numerical upper bound is available. The estimator is based on the collision statistic of a batch of independent samples, a standard unbiased estimator for integer power sums~\cite{AOST}. Its quantum implementation combines a variance estimate with a reflection about the batch state. The latter operation is cheaper than preparing the whole batch again, so we keep the two costs separate. Throughout this section, we write $P=P_k(p)$.

\begin{lemma}[Product-state reflection]\label{lem:product-reflection}
For an integer $r\ge1$ and $0<\zeta<1/2$, the reflection
\[
 2\bigl(\ket{\Psi_p}\bra{\Psi_p}\bigr)^{\otimes r}-I
\]
can be implemented with Euclidean error at most $\zeta$, uniformly on sample-register inputs with initialized auxiliary registers, using $O(\sqrt r\log(1/\zeta))$ queries. A controlled version has the same query bound. The ideal operation returns the auxiliary registers to zero.
\end{lemma}
\begin{proof}
Let $P_j=\ket{\Psi_p}\bra{\Psi_p}$ on register $j$ and $R_j=2P_j-I$. Put
\[
 \ket{u}=r^{-1/2}\sum_{j=1}^r\ket j,
 \qquad S=\sum_{j=1}^r\ket j\bra j\otimes R_j,
 \qquad W=(2\ket u\bra u-I)S.
\]
By routing the selected register through the oracle, we implement $S$ with two queries. Since the projectors $P_j$ commute, they admit a common eigenspace decomposition. On a common eigenspace with exactly $h$ rejecting registers, the eigenphases of $W$ that overlap $\ket u$ are $\pm2\arcsin\sqrt{h/r}$. For $h=0$ the phase is zero; otherwise its distance from zero modulo $2\pi$ is at least $2/\sqrt r$.

Coherent phase detection at resolution $\Theta(r^{-1/2})$, with error $O(\zeta^2)$ and subsequent uncomputation, distinguishes these cases using $O(\sqrt r\log(1/\zeta))$ controlled walk steps~\cite{BHMT}. Applying phase $+1$ in the first case and phase $-1$ in the second implements the desired reflection, because the common positive subspace is spanned by $\ket{\Psi_p}^{\otimes r}$. Controlled local reflections use the known middle phase between ordinary inverse and forward preparation calls, as described in Section~\ref{sec:model}.
\end{proof}

\begin{lemma}[Separate preparation and reflection costs]\label{lem:mean-separate}
Suppose a known observable measured in $\ket\Psi$ produces a real random variable $Z$ with $\Var Z\le\sigma^2$. Suppose preparation costs $C_{\rm prep}$ queries and reflection about $\Psi$ costs $\Oh(C_{\rm ref})$, with logarithmic dependence on inverse implementation error. If the observable is reversibly computable without queries, its mean can be estimated to additive error $\Delta$ using
\begin{equation}\label{eq:mean-separate}
 \Oh\!\left(C_{\rm prep}+C_{\rm ref}\max\{1,\sigma/\Delta\}\right)
\end{equation}
queries at constant success probability. Failure probability $\delta$ introduces an additional factor $O(\log(1/\delta))$.
\end{lemma}
\begin{proof}
The variance-dependent estimator of Montanaro~\cite[Theorem 2.5]{Montanaro} uses a polylogarithmic number of initial-state copies and $\Oh(\sigma/\Delta)$ reflections. Replacing $\sigma$ by $\max\{\sigma,\Delta\}$ covers the remaining case. Its observable computations, shifts, truncations, and rotations are known operations. If the complete procedure makes at most $M$ reflections, implement each with Euclidean error at most $c/M$ for a sufficiently small constant $c$. Telescoping the circuit errors changes the final state by at most $c$. The logarithmic precision overhead and median amplification give the stated bound.
\end{proof}

For independent $X_1,\ldots,X_r\sim p$, $r\ge2k$, define the collision statistic
\begin{equation}\label{eq:statistic}
 Z_r=\binom{r}{k}^{-1}\sum_{\substack{A\subseteq[r]\\|A|=k}}
 \ind\{X_i\text{ is the same for every }i\in A\}.
\end{equation}
This rational observable can be computed reversibly from the sample labels without further queries.

\begin{lemma}[Collision variance]\label{lem:variance}
For fixed $k$ and $r\ge2k$,
\begin{equation}\label{eq:variance}
 \E Z_r=P,\qquad
 \Var Z_r\le C_k\sum_{\ell=1}^k r^{-\ell}P_{2k-\ell}(p)
 \le C_k\sum_{\ell=1}^k r^{-\ell}P^{2-\ell/k}.
\end{equation}
\end{lemma}
\begin{proof}
Each indicator has expectation $P$. Indicators indexed by disjoint subsets are independent. If two subsets of size $k$ overlap in $\ell\ge1$ positions, their product has expectation $P_{2k-\ell}(p)$. For a fixed first subset, there are $\binom{k}{\ell}\binom{r-k}{k-\ell}$ possible second subsets with this overlap. After division by $\binom{r}{k}$, the coefficient is at most $C_k r^{-\ell}$. Summing over $\ell$ gives the first variance estimate, and Eq.~\eqref{eq:highmoment} gives the second.
\end{proof}

\begin{theorem}[Batch estimator with a known upper bound]\label{thm:batch}
For a fixed integer $k\ge2$, suppose $P_k(p)\le R\le1$, where $R$ is known. For any $\eta\in(0,1/16]$ and $\delta\in(0,1/3)$, there exists a quantum algorithm that estimates $P_k(p)$ to within additive error $\eta R$, with success probability at least $1-\delta$, using
\begin{equation}\label{eq:batch}
 \Oh_k\!\left(R^{-1/(k+1)}\eta^{-2/(k+1)}
       +R^{-1/(2k)}\eta^{-1}\right)
\end{equation}
queries, with $\log(1/\delta)$ included among the suppressed factors.
\end{theorem}
\begin{proof}
Choose
\[
 r=\max\{2k,\lceil R^{-1/(k+1)}\eta^{-2/(k+1)}\rceil\},
 \qquad \sigma_R^2=C_k\sum_{\ell=1}^k r^{-\ell}R^{2-\ell/k}.
\]
Preparing the batch requires $r$ queries, whereas Lemma~\ref{lem:product-reflection} implements its reflection using $\Oh(\sqrt r)$ queries. To include the work registers that store the statistic, we add their zero-state projector to the initial reflection and conjugate by the reversible circuit computing $Z_r$. Lemma~\ref{lem:mean-separate} therefore gives
\begin{equation}\label{eq:batchbalance}
 \Oh_k\!\left(r+\frac{\sqrt r\,\sigma_R}{\eta R}\right)
 \le \Oh_k\!\left[r+\frac1\eta
       \left(R^{-1/(2k)}+r^{-(k-1)/2}R^{-1/2}\right)\right].
\end{equation}
To obtain the inequality, observe that the squared terms in the reflection cost form the geometric progression $r^{1-\ell}R^{-\ell/k}$. Its sum is at most $k$ times the sum of its two endpoints. Our choice of $r$ balances the preparation cost with the last term in Eq.~\eqref{eq:batchbalance}. The query budget and the accuracy of each reflection are fixed before running the estimator.
\end{proof}

\medskip
\section{Obtaining a power-sum estimate without prior knowledge}\label{sec:scale}
To estimate the entropy to the required accuracy, we need a constant-factor upper bound on $P_k(p)$. We first show how to query an implicit table of independent samples. We then use a collision-table estimator inside a binary search for the power sum. The proof of the collision-table estimator is deferred to \ref{app:sparse}.

\begin{lemma}[Implicit independent-sample table]\label{lem:table}
An algorithm accessing a length-$L$ table only by
\[
 O_X\ket{j,z}=\ket{j,z\oplus X_j}
\]
can be simulated so that its output law is exactly the law averaged over independent $X_1,\ldots,X_L\sim p$. Each table query, including an inverse or controlled query, uses one call each to $U_p$ and $U_p^\dagger$. The simulation uses $L$ private preparation registers initialized to zero.
\end{lemma}
\begin{proof}
Let $C_j$ copy the visible label in private register $j$ into the response register by XOR. Implement the lookup as
\begin{equation}\label{eq:table-oracle}
 \widetilde O=\sum_j\ket j\bra j\otimes U_{p,j}^\dagger C_jU_{p,j}.
\end{equation}
Routing the addressed register through the oracle port costs one query in each direction. For analysis, set $V=U_p^{\otimes L}$ and $O_{\rm tab}=\sum_j\ket j\bra j\otimes C_j$. Then $\widetilde O=V^\dagger O_{\rm tab}V$. All other algorithmic gates act outside the private table and commute with $V$, so these conjugations telescope. Tracing out the private table removes the final $V^\dagger$.

The resulting computation is equivalent to querying $O_{\rm tab}$ on $\ket{\Psi_p}^{\otimes L}$. Dephasing its label registers changes no output probabilities, because the label projectors commute with the entire computation. The dephased labels have joint law $p^{\otimes L}$. For a controlled query, control only $C_j$; the ordinary preparation calls cancel when the control is inactive. The unitary $V$ is introduced only to analyze the output distribution; the simulation does not apply it and therefore incurs no $L$-query initialization cost.
\end{proof}

\begin{proposition}[Sparse-table estimate]\label{prop:sparse}
For any fixed integer $k\ge2$, suppose $P_k(p)\le g\le1$, where $g$ is known. For any $\eta\in(0,1/4]$ and $\delta\in(0,1/3)$, there is a quantum algorithm that outputs an estimate $\widetilde P_k(p)$ satisfying
\[
 |\widetilde P_k(p)-P_k(p)|\le\eta g
\]
with probability at least $1-\delta$, using $\Oh_k(n^{u_k}/\eta)$ queries. The query count and output range are bounded on every execution, even if the supplied value $g$ fails to satisfy the stated assumption.
\end{proposition}
The proof in \ref{app:sparse} uses a table of length $\lceil k g^{-1/k}\rceil$, whose expected number of $k$-collisions is bounded by a constant depending on $k$. A capped collision count has negligible bias and can be estimated coherently through Lemma~\ref{lem:table}.

\begin{lemma}[Initial scale]\label{lem:scale}
A scale $R$ satisfying
\begin{equation}\label{eq:scale}
 P_k(p)\le R\le\tfrac{16}{3}P_k(p),\qquad R\le1,
\end{equation}
can be found with probability at least $1-\delta$, for any $\delta\in(0,1/3)$, using $\Oh_k(n^{u_k})$ queries. The procedure has a deterministic worst-case query bound.
\end{lemma}

\begin{algorithmblock}{Binary search for the power sum}\label{alg:scale}
\textbf{Input:} Purified access oracle $U_p$; dimension bound $n$; fixed integer $k\ge2$; maximum failure probability $\delta\in(0,1/3)$.\par
\textbf{Parameters:} $J_0:=\lceil(k-1)\log_2n\rceil+2$.\par
\textbf{Output:} A scale $R$ satisfying Eq.~\eqref{eq:scale}.
\begin{enumerate}
\item Set $g\leftarrow1$.
\item For $j=0,1,\ldots,J_0-1$, estimate $P_k(p)$ to additive error $g/16$ using Proposition~\ref{prop:sparse}, with failure probability at most $\delta/(J_0+1)$. Denote the output by $\widetilde P_j$.
\item If $\widetilde P_j\ge g/4$, return $R=g$; otherwise set $g\leftarrow g/2$ and continue.
\item If the iteration limit is reached, declare failure.
\end{enumerate}
\end{algorithmblock}

\begin{proof}
The pseudocode is given in Algorithm~\ref{alg:scale}. Suppose first that every estimate is within its stated error. The initial upper bound $P_k(p)\le1$ is valid. Whenever the algorithm continues, we have $P_k(p)<5g/16<g/2$, so the next call also has a valid upper bound. When it stops, the returned value satisfies $3R/16\le P_k(p)\le R$, as required.

When all estimates are accurate, the procedure stops before the candidate upper bound falls below $P_k(p)$. Since $P_k(p)\ge n^{1-k}$, the prescribed number of iterations is sufficient. By a union bound, all required estimates are accurate with probability at least $1-\delta$. Each test uses $\Oh_k(n^{u_k})$ queries, and there are $O_k(\log n)$ tests. The fixed limit also ensures $g^{-1}=O_k(n^{k-1})$ on unsuccessful histories, so the same query bound holds for every execution.
\end{proof}

The same scale-search procedure can be applied to the batch estimator in Theorem~\ref{thm:batch}. This gives a direct entropy estimator with query complexity
\begin{equation}\label{eq:batchdimension}
 Q_k(n,\eps)\le\Oh_k\!\left(
 n^{(k-1)/(k+1)}\eps^{-2/(k+1)}+n^{b_k}/\eps\right).
\end{equation}
In the full algorithm, we use the batch estimator after a scale has already been obtained.

\medskip
\section{Recursive construction of collision states}\label{sec:prefix}
We next construct the operation that extends a collision state by one order. For convenience, write $\ket{u_i}=\ket{\phi_i}_B\ket i_C$ and $\ket{\Psi_p}=\sum_i\sqrt{p_i}\ket{u_i}$. An order-$s$ collision state consists of $s$ complete purified-sample registers with a common label. When such a state is used to prepare one of higher order, we call it a collision prefix. We use the term cache for a collection of retained copies that will be reused during the construction. A raw sample means a copy of $\ket{\Psi_p}$ before conditioning on any active set. During one stage, the active set is fixed and all operations move complete collision prefixes, including their purification registers.

\subsection{Transduction costs and composition}
We use the transducer formalism of~\cite{BJY} in the form developed in~\cite{CAG}. In particular, we invoke the reflection characterization, rebalancing, amplitude amplification, bounded mean estimation, and composition results in~\cite[Theorem 2.4, Lemma 2.5, Corollary 3.4, Theorem 3.7, and Theorems 2.9--2.15]{CAG}. The underlying state-conversion viewpoint is described in~\cite{LMRSS}. A canonical transducer performs an exact public-state conversion. Its query cost $W(T,v)$ is a quadratic form in the public input vector $v$; hence $W(T,v)^{1/2}$ obeys the triangle inequality. Costs add under sequential composition and direct sums. We use the bidirectional model, permitting $U_p$ and $U_p^\dagger$.

All operations within a complete stage are composed before conversion to a query circuit. A transduction cost $Q$ gives an $O(Q)$-query circuit at constant final accuracy. If stages are converted separately, they use fresh caches and inverse-polylogarithmic accuracy; this overhead is absorbed in $\Oh_k$. In particular, each separately converted stage prepares its own cache, so its exact-input assumption does not depend on the approximation error of a previous conversion.

For a reflection, we refer to the costs on its positive and negative eigenspaces as its positive and negative costs, respectively. Rebalancing changes costs $A,B$ to $aA,B/a$ for any input-independent $a>0$. A useful consequence is the following commuting-projector rule: if $m$ commuting local projectors have reflection cost at most $d$, reflection about their common positive subspace has positive cost $a$ and negative cost $O(md^2/a)$.

For completeness, we give the witness calculation that yields this rule. Let $Q_j$ be the local negative projectors and $K=\sum_jQ_j$. On a vector $v$ orthogonal to their common kernel, use components $Q_jK^+v$, where $K^+$ is the Moore--Penrose inverse. Then
\[
 \sum_jQ_j(Q_jK^+v)=v,
 \qquad \sum_j\norm{Q_jK^+v}^2=\bra vK^+\ket v\le\norm v^2.
\]
The nonzero eigenvalues of $K$ are positive integers. The positive witness has one copy of the input in each index sector, so its squared norm is $m$. These witnesses satisfy the cross-sign reflection constraints. Rebalancing and composition with the local reflections give the rule. The same argument applies when the registers are entangled with untouched auxiliary systems, or when the reflection includes known tests that work registers are zero.

We also record the amplification formula used below. Suppose
\[
 \ket\Psi=\ket{\Psi_0}\ket0+\ket{\Psi_1}\ket1,
 \quad \lambda=\norm{\Psi_1}^2,
 \quad
 \ket{\overline\Psi}=\sqrt{\frac\lambda{1-\lambda}}\ket{\Psi_0}\ket0
 -\sqrt{\frac{1-\lambda}\lambda}\ket{\Psi_1}\ket1.
\]
For a reflection transducer $T$ about $\Psi$, exact preparation of the normalized good state costs
\begin{equation}\label{eq:amplification}
 O\!\left(\frac{W(T,\Psi)+(1-\lambda)W(T,\overline\Psi)}{\sqrt\lambda}\right).
\end{equation}
Reflection about that state has positive cost $O(W(T,\Psi)/\lambda)$ and negative cost $O(W(T,v\ket1))$ on vectors orthogonal to it~\cite[Corollary 3.4]{CAG}.

\subsection{An asymmetric source reflection}
Let $Q$ project onto a cache of $L$ identical order-$s$ prefix states, and let $P_f=\ket{\Psi_p}\bra{\Psi_p}$ act on a fresh register. Suppose a local prefix reflection has cost $D_s$ and restores its initialized retained buffers. The commuting-projector rule gives a cache reflection with costs $b,O(LD_s^2/b)$. A fresh-state reflection can be rebalanced to costs $c,O(1/c)$.

Reflection about $QP_f$ has positive cost $O(b+c)$. On a negative input $z$ its cost is at most
\begin{equation}\label{eq:asymmetric}
 O\!\left(\frac{LD_s^2}{b}\norm{P_fz}^2
       +\frac1c\norm{(I-P_f)z}^2\right).
\end{equation}
To see this, decompose $z=P_fz+(I-P_f)z$. The first component is negative for $Q$, because $QP_fz=0$, and the second is negative for $P_f$. We place the corresponding negative witnesses in separate direct-sum sectors. Their cross-sign constraints sum to the required inner product, and their squared norms give Eq.~\eqref{eq:asymmetric}. This calculation composes the witnesses directly and does not require a separate test of $P_f$. The decomposition is made with the projectors of the negative input, so the local constraints apply across distinct oracle inputs as required. Known index and workspace projectors reject orthogonal components at zero query cost. Previously initialized buffers remain unchanged throughout the operation and are restored exactly.

\subsection{Extending a collision state}
Let $\Act\subseteq\Src$ be known sets and suppose $p_i\le\varphi\le1$ for $i\in\Act$. For an integer $s\ge1$, put
\begin{align*}
 F&=\sum_{i\in\Src}p_i^s>0,&G&=\sum_{i\in\Act}p_i^{s+1},\\
 \ket{\chi_{s,\Src}}&=F^{-1/2}\sum_{i\in\Src}p_i^{s/2}\ket{u_i}^{\otimes s},&
 \ket{\chi_{s+1,\Act}}&=G^{-1/2}\sum_{i\in\Act}p_i^{(s+1)/2}\ket{u_i}^{\otimes(s+1)}.
\end{align*}
The latter state is used only when $G>0$. Write $M_{j,\Act}=\sum_{i\in\Act}p_i^j$.

\begin{lemma}[Exact prefix extension]\label{lem:prefix}
Suppose preparation and reflection for $\chi_{s,\Src}$ have transduction cost at most $D_s$ and restore their initialized buffers. For $L\ge1$, a cache of $L-1$ retained copies, one additional prefix, and a fresh raw sample give preparation of $\chi_{s+1,\Act}$ at cost
\begin{equation}\label{eq:prefixcost}
 D_{s+1}=O\!\left(\sqrt{\frac{D_s^2F\varphi+F/L+\varphi^s}{G}}\right).
\end{equation}
This includes replenishment of the additional prefix. The same bound gives reflection about the new state on $\spanop\{\ket{u_i}^{\otimes(s+1)}:i\in\Act,\ p_i>0\}$. Unused prefixes and all preceding buffers factor exactly from the output and are restored. The guarantees hold with arbitrary spectator entanglement.

The source cache also estimates $G$ to additive error $0<\Delta\le cG$ at additional cost
\begin{equation}\label{eq:prefixmean}
 \Oh\!\left(
 \frac{D_s\sqrt{F\varphi G}+\sqrt{FG/L}+\varphi^{s/2}\sqrt G}{\Delta}
 \right).
\end{equation}
If instead a known upper bound $0\le G\le g\le F\varphi$ is known, $g>0$, and $0<\Delta\le cg$, the bound is
\begin{equation}\label{eq:uppermean}
 \Oh\!\left(
 \frac{D_s\sqrt{F\varphi g}+\sqrt{Fg/L}+\varphi^{s/2}\sqrt g}{\Delta}
 \right).
\end{equation}
This includes $G=0$. Constant-factor estimates suffice to choose the parameters, while the final numerical rescaling requires a correspondingly accurate estimate of $F$. Mean estimation restores the source cache and the fresh sample exactly at the transducer level.
\end{lemma}

\begin{proof}
Choose a uniform index in $[L]$, compare the selected prefix label with a fresh raw sample, and flag a match in $\Act$. Move the selected complete prefix to an output port. Let $V$ denote these query-free operations. The good component factors as
\begin{equation}\label{eq:factor}
 \sqrt h\,\ket{\chi_{s+1,\Act}}
 \otimes\ket{\chi_{s,\Src}}^{\otimes(L-1)}\otimes\ket{u_L},
 \qquad h=G/F,
\end{equation}
where $\ket{u_L}$ is a known uniform index state. We uncompute this known index state after the operation. We also include an independent coin with success probability $\vartheta=1/4$ in the flag. The actual success probability is $\lambda=\vartheta h\le1/4$, with the same normalized good state.

First consider a unit vector $v=\sum_{i\in\Act}v_i\ket{u_i}^{\otimes(s+1)}$ orthogonal to the new prefix. Invert the witness map on $v$ with the retained prefixes and uniform index, and project the initial known workspace onto its prescribed state. Call the resulting vector $B_v$ for the undiluted flag; the actual flag gives $\sqrt\vartheta B_v$. With $V_j=\sum_{i\in\Act}p_i^j|v_i|^2$, direct inner products give
\begin{equation}\label{eq:negativeidentities}
 \norm{B_v}^2=\frac1L+\left(1-\frac1L\right)\frac{V_s}{F},
 \quad \norm{P_fB_v}^2=\frac{V_1}{L},
 \quad QP_fB_v=0.
\end{equation}
Equal insertion positions contribute $1$ to the first identity; distinct positions contribute $V_s/F$. After projection of the fresh register onto $\Psi_p$, the inserted prefix vector is $\sum_i\sqrt{p_i}v_i\ket{u_i}^{\otimes s}$. Its overlap with the cache state is zero by the orthogonality of $v$ to $\chi_{s+1,\Act}$. All cross-position terms therefore vanish in the second identity. Allowing the coefficients $v_i$ to take values in an auxiliary Hilbert space proves the same identities in the presence of spectator entanglement.

Combining Eqs.~\eqref{eq:asymmetric} and~\eqref{eq:negativeidentities} with the reflection formula for the normalized good state gives positive and negative costs bounded by
\[
 O\!\left(\frac{b+c}{\vartheta h}\right),
 \qquad
 O\!\left(\vartheta\frac{D_s^2\varphi}{b}
     +\vartheta\frac{1/L+\varphi^s/F}{c}\right),
\]
respectively. We set $b=c$ and choose their common value to balance these two bounds. The squared balanced cost is a constant multiple of $(D_s^2\varphi+1/L+\varphi^s/F)/h$, proving the reflection part of~\eqref{eq:prefixcost}.

For preparation and estimation, let $Z_1,\ldots,Z_L$ be independent cache labels with law $q_i=p_i^s/F$ on $\Src$. Let $N_i$ count their occurrences of $i$, and for an independent raw label $Y\sim p$ put
\[
 X=\frac{N_Y\ind_{Y\in\Act}}L,
 \qquad S=\sum_{a=1}^L p_{Z_a}\ind_{Z_a\in\Act}.
\]
Binomial moments give
\begin{align}
 \E X&=h,&
 \Var X&=\frac hL+\left(1-\frac1L\right)\frac{M_{2s+1,\Act}}{F^2}-h^2,\label{eq:varX}\\
 \E S&=Lh,&
 \Var S&=L\left(\frac{M_{s+2,\Act}}F-h^2\right).\label{eq:varS}
\end{align}
Let $\overline\Psi$ be the normalized Grover complement of the flagged source. After inverse witness mapping and projection of the known workspace, its coefficient in each cache and fresh-label sector is $\vartheta(h-X)/\sqrt{\lambda(1-\lambda)}$. For the resulting vector $B_{\overline\Psi}$,
\begin{align}
 \norm{B_{\overline\Psi}}^2
 &=\frac{\vartheta^2\Var X}{\lambda(1-\lambda)}
 \le O\!\left(\frac1L+\frac{M_{2s+1,\Act}}{FG}\right),\label{eq:complement1}\\
 \norm{P_fB_{\overline\Psi}}^2
 &=\frac{\vartheta^2\Var S}{L^2\lambda(1-\lambda)}
 \le O\!\left(\frac{M_{s+2,\Act}}{LG}\right),
 \qquad QP_fB_{\overline\Psi}=0.\label{eq:complement2}
\end{align}
The source reflection, conjugated by $V$, consequently has costs
\begin{align*}
 W(T,\Psi)&=O(b+c),\\
 W(T,\overline\Psi)&=O\!\left(
 \frac{D_s^2M_{s+2,\Act}}{bG}
 +\frac{1/L+M_{2s+1,\Act}/(FG)}c\right).
\end{align*}
Use $M_{s+2,\Act}\le\varphi G$ and $M_{2s+1,\Act}\le\varphi^sG$, and choose
\[
 b=c=\sqrt{D_s^2\varphi+1/L+\varphi^s/F}.
\]
The sum of the two costs is $O(\sqrt{D_s^2\varphi+1/L+\varphi^s/F})$. Eq.~\eqref{eq:amplification} gives Eq.~\eqref{eq:prefixcost} for preparation. The same bound includes the cost of replacing the consumed prefix, since $G\le F\varphi$. It also includes the fresh sample, since $G\le\varphi^s\sum_{i\in\Act}p_i\le\varphi^s$. The factorization in Eq.~\eqref{eq:factor} is exact.

For the mean, apply the bounded-mean transducer of~\cite[Theorem 3.7]{CAG} to the Boolean flag observable. Request flag error $\Delta_\lambda=\vartheta\Delta/F$ and choose its iteration parameter $t=\Theta(\sqrt\lambda/\Delta_\lambda)$. Choose its bias and variance parameter at most a sufficiently small multiple of $\Delta_\lambda^2$. The output variance is $O(\lambda/t^2)+O(\Delta_\lambda^2)$, while the cost is, up to logarithms, $t$ times the sum of the source and complement costs. A constant number of repetitions and Chebyshev's inequality give Eq.~\eqref{eq:prefixmean}. The source is restored exactly; uncomputing $V$ restores its cache and fresh sample. Only the source reflection and the known flag enter this calculation. Hence the bound is independent of the purification vectors and their phases.

To prove the statement with a known upper bound $g$, choose a known $\rho=\Theta(g/F)$ with $0<\rho\le1/8$ and a sufficiently small fixed $\vartheta>0$. A three-branch known coin always accepts with weight $\rho$, accepts an active match with weight $\vartheta$, and otherwise rejects. Its mean is
\[
 \mu=\rho+\vartheta G/F=\Theta(g/F),\qquad 0<\mu\le1/2.
\]
The centered coefficient after inverse witness mapping is $\vartheta(h-X)/\sqrt{\mu(1-\mu)}$. The squared norms are $\vartheta^2\Var X/[\mu(1-\mu)]$ and $\vartheta^2\Var S/[L^2\mu(1-\mu)]$. The moment contributions in Eq.~\eqref{eq:complement1}--\eqref{eq:complement2} now have an additional factor at most $G/g$. They vanish when $G=0$. We use the same reflection-cost parameters and take $t=\Theta(\sqrt{g/F}/(\Delta/F))$. Subtracting the known baseline and multiplying by $F/\vartheta$ gives Eq.~\eqref{eq:uppermean}. This procedure remains well defined when $G=0$, because it does not require preparation of the corresponding normalized collision state.
\end{proof}

\subsection{Order-two states and retained auxiliary registers}
Lemma~\ref{lem:prefix} includes $s=1$. Take the unconditioned source on the full alphabet, so $\Src=[n]$, $M_1=1$, $D_1=O(1)$, and $\chi_{1,\Src}=\Psi_p$. With a raw cache of size $r$, the active pair state satisfies
\begin{equation}\label{eq:pairbase}
 D_2^2=O\!\left(\frac{1/r+\varphi}{M_2}\right),
 \qquad M_2=\sum_{i\in\Act}p_i^2.
\end{equation}
There is no need to condition this source on the active set. The mean estimator with a known upper bound estimates $M_2$ before the normalized pair state is constructed, and it remains well defined when $M_2=0$.

Local reflections preserve their initialized buffers. On that public subspace, the local projectors of a cache commute even when their implementations share earlier buffers. We may therefore compose their reflection witnesses with the exact transducers from the preceding levels. The construction uses only their action on this invariant subspace.

A stage with $L_s$ initialized order-$s$ prefixes and $M$ order-$k$ outputs needs $O_k(\sum_s sL_s+k{}M+1)$ raw samples, in addition to its stated raw buffer. We prepare this sample pool in advance. Subsequent controlled operations use the prepared samples, known gates, and controlled reflections, without controlled calls to $U_p$. Every prefix cost is at least a constant, so the pool cost is included in setup and per-output costs. Early termination leaves the unused pool untouched. A complete stage is composed before circuit conversion, with fresh pools at any separately converted stage.

\medskip
\section{Query complexity of the recursive construction}\label{sec:budgets}
In this section, we analyze the number of retained states required by the recursive construction. Fix an integer $k\ge2$ and suppose that a known scale satisfies $P_k(p)\le R\le C_k P_k(p)$ and $R\le1$. Let $\eta$ denote the relative power-sum accuracy. Constants depending only on $k$ are absorbed into the choices of the parameters.

\subsection{Choice of accuracy regime}
If $\eta\le R^{1/(2k)}$, Theorem~\ref{thm:batch} gives
\begin{equation}\label{eq:highprecision}
 \Oh_k\!\left(R^{-1/(k+1)}\eta^{-2/(k+1)}
       +R^{-1/(2k)}/\eta\right)
 =\Oh_k(R^{-1/(2k)}/\eta).
\end{equation}
We therefore analyze the branch
\begin{equation}\label{eq:recursivebranch}
 \eta>R^{1/(2k)},\qquad \varphi_0=R^{1/k},\qquad
 \varphi_*=(\eta R)^{1/k}.
\end{equation}
The stage threshold satisfies $\varphi_*\le\varphi\le \varphi_0$. In particular,
\begin{equation}\label{eq:thresholdbound}
 \varphi\ge R^{1/(k-1)},
\end{equation}
since $\eta>R^{1/(2k)}\ge R^{1/(k-1)}$. Choose
\begin{equation}\label{eq:budget}
 B=R^{-u_k/(k-1)}\eta^{-(1-u_k)},\qquad r=\lceil B\rceil.
\end{equation}
When implementing the algorithm, we multiply these budgets by sufficiently large constants depending only on $k$.

\subsection{Initial states and the recursive query bound}
The exponents satisfy
\begin{equation}\label{eq:uelementary}
 u_{k,j}\le\frac{k-1}{k+1}\qquad(2\le j\le k).
\end{equation}
Indeed, with $M=2^{j-2}$, the numerator of the difference after multiplication by the positive common denominator is
\[
 M\bigl[k^2-7k+j(k+1)\bigr]+2k.
\]
For $j=2$ it is $(k-1)(k-2)$. For $j\ge3$, the bracket is at least $(k-1)(k-3)\ge0$. It follows from Eq.~\eqref{eq:recursivebranch} that
\begin{equation}\label{eq:BP}
 B\varphi_0\le R^{\{(k-1)-(k+1)u_k\}/\{2k(k-1)\}}\le1.
\end{equation}
At a fixed active set, write $M_s=\sum_{i\in\Act}p_i^s$ for $s\ge2$, and keep $M_1=1$ for the unconditioned source. By~\eqref{eq:interpolation},
\begin{equation}\label{eq:Hsbound}
 M_s\le R^{(s-1)/(k-1)}\qquad(2\le s\le k).
\end{equation}
The pair base and Eq.~\eqref{eq:BP} give
\begin{equation}\label{eq:Ebase}
 E_2^2:=M_2D_2^2=O_k(2\varphi+B^{-1})=O_k(B^{-1}).
\end{equation}

Once a constant-factor estimate $\widetilde M_s=\Theta_k(M_s)$ has been certified, choose a bound $D_s\ge1$ on the local preparation and reflection costs and retain
\begin{equation}\label{eq:cachesize}
 L_s=\max\{1,\lceil B/D_s\rceil\},\qquad 2\le s<k.
\end{equation}
The next query-cost estimate is computed from the exact-prefix numerator and a certified scale for $M_{s+1}$:
\begin{equation}\label{eq:localproxy}
 D_{s+1}=C_k\max\left\{1,
 \sqrt{\frac{D_s^2\widetilde M_s\varphi+\widetilde M_s/L_s+\varphi^s}
                  {\widetilde M_{s+1}}}\right\}.
\end{equation}
As shown in Section~\ref{sec:algorithm}, the numerator can be computed before obtaining a constant-factor estimate of $M_{s+1}$. We use this local recurrence to determine the number of retained copies at each order.

Set $E_s^2=M_sD_s^2$. Since $L_s\ge B/D_s$, the recurrence implies
\begin{equation}\label{eq:Erec}
 E_2^2=O_k(B^{-1}),\qquad
 E_{s+1}^2\le C_k\left(\varphi E_s^2+
        \frac{\sqrt{M_s}\,E_s}{B}+\varphi^s\right).
\end{equation}
The contribution of the maximum with $1$ in Eq.~\eqref{eq:localproxy} is absorbed by the inequality $M_{s+1}\le\varphi^s$. Rounding $L_s$ upwards can only decrease the middle term. Thus Eq.~\eqref{eq:Erec} holds independently of the setup-cost bound that we prove below.

\subsection{Analysis of the recurrence}
Normalize by $y_s=E_s^2/\varphi^{s-1}$. Eqs.~\eqref{eq:Hsbound}--\eqref{eq:Erec} give
\begin{equation}\label{eq:yrec}
 y_2=O_k((B\varphi)^{-1}),\qquad
 y_{s+1}\le C_k(y_s+a_s\sqrt{y_s}+1),\qquad
 a_s=\frac{R^{(s-1)/(2(k-1))}}{B\varphi^{(s+1)/2}}.
\end{equation}
The coefficients decrease with $s$, because
\[
 \frac{a_{s+1}}{a_s}=\sqrt{R^{1/(k-1)}/\varphi}\le1.
\]
Moreover, $B\varphi\le1$. We may therefore absorb the additive constant into a fixed multiple of the initial bound $(B\varphi)^{-1}$.

To expand the recurrence, bound each sum by a fixed multiple of its largest summand. Every resulting term is obtained by composing identity maps with maps of the form $z\mapsto a_s\sqrt z$. If it uses $m$ square-root steps at indices $i_1<\cdots<i_m$, then $i_j\ge j+1$. Because $a_s$ is nonincreasing, moving them to indices $2,\ldots,m+1$ can only enlarge that term. Thus
\begin{equation}\label{eq:path}
 y_s\le C_k'\max_{0\le m\le s-2}V_m,
 \qquad
 V_m=R^{(m-1+2^{-m})/(k-1)}
 B^{-(2-2^{-m})}\varphi^{-(m+1)}.
\end{equation}
For $m=0$ this is exactly $V_0=(B\varphi)^{-1}$.

We show that each term is at most
\begin{equation}\label{eq:Y}
 Y=\frac{B^2\varphi^{k+1}}{R}.
\end{equation}
Put $j=m+2$ and $M=2^m$. The inequality $V_m\le Y$ is equivalent to
\begin{equation}\label{eq:pathcondition}
 B^{4M-1}\ge
 R^{\{M(k+j-4)+1\}/(k-1)}\varphi^{-M(k+j)}.
\end{equation}
The right-hand side is largest at $\varphi=\varphi_*$. Substitution and a $(4M-1)$st root yield
\begin{equation}\label{eq:Bj}
 B\ge B_j:=R^{-u_{k,j}/(k-1)}\eta^{-(1-u_{k,j})}.
\end{equation}
The identity $1-u_{k,j}=2^{j-2}(k+j)/[k(2^j-1)]$ verifies the accuracy exponent directly. Moreover,
\[
 B_j=\eta^{-1}(\eta R^{-1/(k-1)})^{u_{k,j}},
\]
and the factor in parentheses is at least one in the regime of Eq.~\eqref{eq:recursivebranch}. Maximizing over $j$ therefore gives precisely the budget in Eq.~\eqref{eq:budget}. Combining Eqs.~\eqref{eq:path}--\eqref{eq:Y}, we obtain the following bound simultaneously for all orders:
\begin{equation}\label{eq:envelope}
 E_s^2\le O_k\!\left(\frac{B^2\varphi^{k+s}}{R}\right),
 \qquad 2\le s\le k.
\end{equation}

\begin{lemma}[Uniform setup budgets]\label{lem:budgets}
In the branch~\eqref{eq:recursivebranch}, the raw cache costs $O_k(B)$. For every retained prefix whose moment satisfies $M_s\gtrsim_k\varphi^s$, the recursive query-cost estimate satisfies $D_s=O_k(B)$. Its rounded cache has setup cost $L_sD_s=O_k(B)$. All retained caches together have cost $O_k(B)$ for fixed $k$, and Eq.~\eqref{eq:envelope} holds for every prefix.
\end{lemma}
\begin{proof}
The simultaneous bound in Eq.~\eqref{eq:envelope} has already been established. For any retained prefix, it gives
\[
 D_s^2=E_s^2/M_s\le O_k(B^2\varphi^k/R)=O_k(B^2),
\]
because $\varphi\le \varphi_0$. Hence $L_sD_s\le B+D_s=O_k(B)$. The raw cache has $\lceil B\rceil$ samples. There are at most $k-2$ additional caches. We construct a prefix only after certifying the stated condition on its moment. Section~\ref{sec:algorithm} describes this test and explains how the algorithm proceeds when the moment is too small.
\end{proof}

At $k=2$ there are no square-root steps and no intermediate collision caches. The only path is $V_0$, and
\[
 B=R^{-1/3}\eta^{-2/3},\qquad
 \varphi_*=(\eta R)^{1/2},\qquad B^3\varphi_*^4/R=1.
\]
Hence the pair-state construction satisfies Eq.~\eqref{eq:envelope} as well. Its source normalization is known exactly, since $M_1=1$.

\medskip
\section{\Renyi{} entropy estimator for integer orders}\label{sec:algorithm}
In this section, we combine the preceding subroutines to estimate $H_k(p)$. We first locate the power sum within a constant factor. We then estimate the contributions of the recorded and active labels separately and analyze the error and query complexity of the complete algorithm.

\subsection{Algorithm and parameter choice}
Let $\eps$ be the required entropy accuracy. We choose $\eta=c'_k\eps$ for a sufficiently small constant $c'_k>0$, decreasing the allowed constant range of $\eps$ if necessary so that $\eta\le1/64$. First obtain $R$ using Algorithm~\ref{alg:scale}. If $\eta\le R^{1/(2k)}$, apply the batch estimator of Theorem~\ref{thm:batch}. Its query complexity is given by Eq.~\eqref{eq:highprecision}. Otherwise, set
\begin{equation}\label{eq:algorithmparameters}
 \Delta=\eta R,\qquad \varphi_*=\Theta_k(\Delta^{1/k}),\qquad
 \varphi_0=R^{1/k},\qquad
 B=C_k{}R^{-u_k/(k-1)}\eta^{-(1-u_k)},\qquad r=\lceil B\rceil.
\end{equation}
We choose the constants in $\varphi_*$ and $B$ to satisfy Lemma~\ref{lem:budgets} and the error requirements below. Let $\varphi_j=\varphi_0/2^j$ and choose the final index so that its threshold is within a fixed factor of $\varphi_*$. The pseudocode is given in Algorithm~\ref{alg:main}.

\begin{algorithmblock}{Quantum algorithm for integer order \Renyi{} entropy estimation}\label{alg:main}
\textbf{Input:} Purified access oracle $U_p$ for $p=(p_i)_{i=1}^n$; fixed integer $k\ge2$; additive accuracy $\eps\in(0,c_k]$; maximum failure probability $\delta\in(0,1/3)$.\par
\textbf{Parameters:} $\eta:=c'_k\eps$; the parameters in Eq.~\eqref{eq:algorithmparameters} after $R$ is found.\par
\textbf{Output:} An estimate $\widetilde H_k(p)$ satisfying Problem~\ref{prob:entropy}.
\begin{enumerate}
\item Use Algorithm~\ref{alg:scale} to find $R$, with failure probability at most $1/20$.
\item If $\eta\le R^{1/(2k)}$, estimate the full moment with the batch estimator and go to step 6. Otherwise initialize the parameters in Eq.~\eqref{eq:algorithmparameters}, set $\Act=[n]$, and set the recorded set to empty.
\item At each dyadic threshold $\varphi$, estimate the active moments $M_2,\ldots,M_k$ within constant factors, in increasing order. Construct a prefix cache only after certifying that its moment is large enough. If a moment is sufficiently small, halve $\varphi$ without collecting labels.
\item At a retained nonfinal stage, keep the active mask fixed and draw $\Oh_k(M_k/\varphi^k)$ labels from the order-$k$ state. Add the observed labels to the recorded set, remove them from activity, and halve $\varphi$.
\item At the final threshold, either certify a negligible active moment or refine the required normalizations and estimate that moment. Estimate the recorded contribution by conditional batch estimation.
\item Denote the sum of the estimated contributions by $\widetilde P$. Project it onto $[n^{1-k},1]$ and output $\log\widetilde P/(1-k)$ for this run.
\item Repeat the complete run $O(\log(1/\delta))$ times independently and return the median as $\widetilde H_k(p)$.
\end{enumerate}
Within a run, use fixed error assignments so that steps 1--6 succeed jointly with probability at least $2/3$. Every separately converted test or stage uses fresh retained states and a predetermined query limit.
\end{algorithmblock}

\subsection{Preparation of the retained states}
At a fixed stage, maintain $p_i\le\varphi$ for $i\in\Act$. Define $M_s=\sum_{i\in\Act}p_i^s$ for $2\le s\le k$ and $G=M_k$, while $M_1=1$ denotes the normalization of the unconditioned source. We use
\begin{equation}\label{eq:activebounds}
 M_s\le R^{(s-1)/(k-1)},\qquad
 M_{s+1}\le\varphi M_s\ (s\ge2),\qquad
 G\le\varphi^{k-s}M_s\ (2\le s\le k).
\end{equation}
The pair moment also satisfies $M_2\le\varphi M_1$.

The unnormalized query-cost bound $e_s$ can be computed before estimating $M_s$. Start with $e_2=C_k(2\varphi+1/r)$. Once a coarse scale $\widetilde M_s=\Theta_k(M_s)$ is certified, set
\begin{equation}\label{eq:explicitproxies}
 \begin{aligned}
 D_s&=C_k\max\{1,\sqrt{e_s/\widetilde M_s}\},\qquad
 L_s=\max\{1,\lceil B/D_s\rceil\},\\
 e_{s+1}&=C_k(D_s^2\widetilde M_s\varphi+\widetilde M_s/L_s+\varphi^s).
 \end{aligned}
\end{equation}
The fixed constants may be chosen separately at successive orders. With this choice, $e_{s+1}$ is known before the next moment is estimated. Taking the maximum with $1$ does not change the asymptotic bound, because $e_s\gtrsim_k\varphi^{s-1}\ge M_s$. An extra source prefix, in addition to the retained cache, is included when applying Lemma~\ref{lem:prefix}; this changes the bounds by only a constant.

By induction, these quantities bound the exact preparation and reflection costs. They also satisfy
\begin{equation}\label{eq:erec}
 \begin{aligned}
 e_2&=O_k(B^{-1}),\\
 e_{s+1}&\le O_k\!\left(\varphi e_s+
       \frac{R^{(s-1)/(2(k-1))}\sqrt{e_s}}B+\varphi^s\right),\\
 e_s&\le O_k(B^2\varphi^{k+s}/R).
 \end{aligned}
\end{equation}
The last inequality follows from the recurrence analysis in Section~\ref{sec:budgets}. It is used only in the analysis; the algorithm determines cache sizes from the computable quantities in Eq.~\eqref{eq:explicitproxies}. All retained buffers are restored exactly. A certified inequality $M_s\gtrsim_k\varphi^s$ implies $D_s=O_k(B)$ and $L_sD_s=O_k(B)$ by Lemma~\ref{lem:budgets}.

\subsection{Estimating the restricted power sums}\label{sec:activelocalization}
We obtain constant-factor estimates of $M_2,M_3,\ldots,M_k$ in increasing order. Each test for $M_s$ uses only caches whose normalizations have already been certified. The first test needs the unconditioned source and its buffer, so it is defined even when the active set has zero mass.

For $M_s$, begin with a valid upper scale derived from $R$ and the coarse interval for $\varphi M_{s-1}$; for $s=2$ use $M_1=1$. We decrease the candidate upper bound geometrically until it reaches $c_s\varphi^s$, where $c_s>0$ is a sufficiently small fixed constant. At a candidate scale $g$, apply~\eqref{eq:uppermean} to additive precision $\Theta_k(g)$. Its query cost, apart from the bounded initialization, is
\begin{equation}\label{eq:localizecost}
 \Oh_k(\sqrt{e_s/g})
 \le\Oh_k(B\sqrt{\varphi^k/R})
 \le\Oh_k(B),\qquad g\gtrsim_k\varphi^s.
\end{equation}

We implement the tests with confidence intervals and stopping thresholds separated by fixed factors. Whenever a test is accurate, it either certifies a smaller valid upper bound or returns a constant-factor interval containing $M_s$. Constant-factor normalization estimates suffice at this point: we multiply the estimated match ratio by the certified endpoints for $M_{s-1}$. Their ratio is bounded in terms of $k$. If the supplied scale is within a fixed factor of $M_{s-1}\varphi$, reduce the fixed coin coefficient in the baseline construction accordingly. This preserves its cost bound.

If the test certifies $M_s<(\varphi/2)^s$, no active label has probability at least $\varphi/2$. Skip extraction and halve $\varphi$. The tests can be continued to a sufficiently small floor so that a successful exit always either yields this certificate or returns $M_s\gtrsim_k\varphi^s$. At the final threshold, a small certificate gives
\[
 G\le\varphi^{k-s}M_s=O_k(\Delta),
\]
with its constant chosen within the residual error allocation. We may then set the residual estimate to zero. Otherwise, the certified moment is large enough to construct the next prefix cache. Thus every normalized state used by the algorithm has a positive certified normalization, while the positive acceptance baseline keeps the preceding tests well defined even at zero.

\subsection{Finding labels and estimating the remaining contribution}
At a retained nonfinal stage, let $G=M_k\gtrsim_k\varphi^k$. Use the certified scale to select
\[
 M=\Oh_k(G/\varphi^k)
\]
independent output labels from the exact order-$k$ state, with the mask fixed for the whole stage. Each active label with $p_i\ge\varphi/2$ is sampled with probability at least $\varphi^k/(2^k{}G)$. A union bound shows that the logarithmic factor in the number of samples suffices to observe all such labels with the required confidence. Remove every observed label at the end of the stage. On this event, every label remaining in the active set has probability at most $\varphi/2$.

By~\eqref{eq:erec}, the extraction cost is
\begin{equation}\label{eq:extractioncost}
 MD_k\le\Oh_k\!\left(\frac{\sqrt{G e_k}}{\varphi^k}\right)
 \le\Oh_k(B\sqrt{G/R})=\Oh_k(B).
\end{equation}
The total number of recorded labels is at most the number of outputs. Summing the geometric threshold bounds gives
\begin{equation}\label{eq:knownsize}
 |S|\le\Oh_k(R/\varphi_*^k)=\Oh_k(1/\eta).
\end{equation}
At the final stage the active probabilities are $O_k(\varphi_*)$.

If the residual has not already been certified negligible, find a constant-factor estimate $g=\Theta_k(G)$ above a sufficiently small multiple of $\Delta$. The prefix mean primitive estimates it to additive error $\Theta_k(\Delta)$ at cost
\begin{equation}\label{eq:residualcost}
 \Oh_k\!\left(B+\frac{\sqrt{e_k g}}\Delta\right)
 \le\Oh_k\!\left(B+B\frac{\varphi^k}\Delta\sqrt{g/R}\right)
 =\Oh_k(B).
\end{equation}
The last equality uses the final threshold relation $\varphi^k=\Theta_k(\Delta)$. The residual is discarded only after its total moment has been certified small. A bound on the largest remaining probability alone would not justify this step.

\subsection{Accurate normalization factors}
The normalized collision states are defined in terms of the true moments, even though the algorithm initially knows them only within constant factors. After undoing known coin factors, their match probabilities obey
\[
 M_s=M_{s-1}\lambda_s,\qquad 2\le s\le k,\qquad M_1=1.
\]
Constant-factor estimates suffice for constructing the states, locating subsequent moments, and collecting labels. For final numerical rescaling, let $g=\Theta_k(G)$ be a retained residual scale and put $\nu=\Theta_k(\min\{1,\Delta/g\})$. Refine $M_2,\ldots,M_{k-1}$ sequentially to relative error $O_k(\nu)$, assigning a sufficiently small fixed fraction to each factor. The accumulated multiplicative error over the fixed number of orders is $O_k(\nu)$. At $k=2$ this list is empty.

For $2\le s<k$, the refinement costs
\[
 \Oh_k\!\left(B+\frac{gD_s}{\Delta}\right)
 \le\Oh_k\!\left(B+\frac{g\sqrt{e_s}}{\Delta\sqrt{M_s}}\right).
\]
Using $G\le\varphi^{k-s}M_s$, $g=\Theta_k(G)$, and Eq.~\eqref{eq:erec},
\begin{equation}\label{eq:calibration}
 \frac{g\sqrt{e_s}}{\Delta\sqrt{M_s}}
 \le O_k\!\left(B\frac{\varphi^k}\Delta\sqrt{g/R}\right)
 =O_k(B).
\end{equation}
The normalization refinement therefore adds only $\Oh_k(B)$ queries. We perform it only after certifying that the residual is large enough to require estimation. Consequently, no relative-accuracy requirement is imposed on a zero residual.

\subsection{Estimating the contribution of the recorded labels}
\begin{lemma}[Estimation on a known set]\label{lem:known}
Let $S\subseteq[n]$ be a known set of $m$ labels, $P_k(p)\le R$, and $\Delta=\eta R$. Its contribution $G_S=\sum_{i\in S}p_i^k$ can be estimated to additive error $O_k(\Delta)$ using
\begin{equation}\label{eq:knowncost}
 \Oh_k\!\left[
 R^{-1/(2k)}\left(
 m^{(k-1)^2/(2k(k+1))}\eta^{-2/(k+1)}+\eta^{-1}
 \right)\right]
\end{equation}
purified queries. If $m=\Oh_k(1/\eta)$, this is $\Oh_k(R^{-1/(2k)}/\eta)$.
\end{lemma}
\begin{proof}
If $S$ is empty, we return zero. Otherwise, put $w=\sum_{i\in S}p_i$ and $F=\min\{R,w^k\}$. Then $G_S\le F$, and H\"older's inequality gives $w^k\le m^{k-1}G_S\le m^{k-1}F$.

Suppose first that a constant-factor scale for positive $w$ is known. Fixed-point amplitude amplification~\cite{YLC} prepares the original purified state conditioned on $S$ at cost $\Oh(w^{-1/2})$. This supplies purified access to the distribution $p_i/w$ on $S$, whose power sum of order $k$ is at most $F/w^k$. Apply Theorem~\ref{thm:batch} with additive precision $\Theta_k(\Delta/w^k)$. If the requested error is already a sufficiently large fraction of the upper bound on the conditional moment, we return zero instead. Including conditional preparation, the query cost is
\begin{equation}\label{eq:knownconditional}
 \Oh_k\!\left[
 w^{(k-1)/(2(k+1))}F^{1/(k+1)}\Delta^{-2/(k+1)}
 +F^{1-1/(2k)}/\Delta\right].
\end{equation}
We enlarge the moment upper bounds by fixed factors when necessary to account for the uncertainty in the mass estimate. Substituting $w^k\le m^{k-1}F$ and $F\le R$ gives Eq.~\eqref{eq:knowncost}. For $m=\Oh_k(1/\eta)$, the first inverse-accuracy exponent is
\[
 \frac{(k-1)^2}{2k(k+1)}+\frac2{k+1}=\frac{k+1}{2k}\le1.
\]

To find the scale of $w$, we use amplitude-estimation tests~\cite{BHMT} with geometrically decreasing candidate values and fixed query budgets, stopping at a sufficiently small multiple of $\Delta^{1/k}$. If $w$ is certified below that scale, then $G_S\le w^k=O_k(\Delta)$ and returning zero is valid. Otherwise retain a constant-factor interval and refine $w$ to relative error $\Theta_k(\min\{1,\Delta/F\})$. The refinement costs
\begin{equation}\label{eq:masscost}
 \Oh_k(F/(\Delta\sqrt w))\le\Oh_k(R^{-1/(2k)}/\eta).
\end{equation}
Indeed, if $w^k\le R$ then $F/\sqrt w=w^{k-1/2}\le R^{1-1/(2k)}$; if $w^k\ge R$, the same conclusion follows from $F=R$. The initial tests cost $\Oh_k(\Delta^{-1/(2k)})$, which is absorbed. The refined mass controls the final multiplication by $w^k$ to error $O_k(\Delta)$, since the contribution is at most $F$. When $w=0$, the initial tests terminate without invoking conditional preparation.

Finally, fix an upper bound $Q$ on the number of conditional preparation and inverse calls in the complete capped batch circuit. Implement each preparation with state error at most
\[
 \xi=c_k\min\{1,\Delta/w^k\}/(Q+1)^2,
\]
where we choose the tolerance conservatively from the certified interval for $w$. A two-dimensional output rotation gives an ideal unitary extension of the exact conditional state whose operator-norm distance from the implemented preparation is $O(\xi)$. Its inverse has the same error, so the full circuit changes by at most $O(Q\xi)$. Fixed-point amplification has only logarithmic dependence on $1/\xi$. The known positive mass floor and $R\ge n^{1-k}$ keep this logarithm polynomial in $\log(n/\eta)$. Allocate fixed fractions of the total error to the mass, conditional moment, and preparation errors, and amplify confidence to obtain the claim.
\end{proof}

\subsection{Error analysis and total query complexity}
We assign a fixed query budget to every scale test, cache construction, and output stage, using the certified classical intervals available at that point. Each complete stage receives a fresh raw-sample pool covering its cache and output budgets. Its preparation is charged through the setup and per-output costs. Controlled reflections about the raw sample use only ordinary $U_p,U_p^\dagger$ and a known controlled zero-state phase.

Within each stage, we compose initialization with all label-extraction operations before converting the transducer to a circuit. The retained source states are therefore exact throughout the transducer construction. There are $O_k(\log^2(n/\eta))$ scale tests and stages. Rebuild caches for every separately converted test or stage, and use inverse-polylogarithmic conversion accuracy. Confidence amplification and a union bound add only logarithmic factors. In particular, extraction calls within a stage are not converted individually at accuracy inverse-polynomial in the number of outputs.

The endpoint intervals fix the maximum number of scale tests. Cap every test, cache construction, and extraction stage by its prescribed $\Oh_k(B)$ bound, and cap the number of recorded labels by Eq.~\eqref{eq:knownsize}. An exceeded cap or an inconsistent confidence interval is declared a failure. With sufficiently large fixed constants in the budgets, these events do not occur when all estimates are accurate. The caps also bound the number of queries on every other execution.

\begin{proof}[Proof of Theorem~\ref{thm:main}]
Obtaining the initial scale requires $\Oh_k(n^{u_k})$ queries. In the high-precision branch,~\eqref{eq:highprecision} costs $\Oh_k(R^{-1/(2k)}/\eta)$. Otherwise each localization, setup, extraction, normalization, and residual step costs $\Oh_k(B)$ by Eq.~\eqref{eq:localizecost}--\eqref{eq:calibration} and Lemma~\ref{lem:budgets}. Lemma~\ref{lem:known} and Eq.~\eqref{eq:knownsize} estimate the recorded contribution at cost $\Oh_k(R^{-1/(2k)}/\eta)$.

Add the estimates for the recorded and active labels. Their total moment error is $O_k(\eta R)=O_k(\eta P_k(p))$. Since $\eta=c'_k\eps$, choosing $c'_k$ sufficiently small and applying Eq.~\eqref{eq:log} gives entropy error at most $\eps$. The projection onto $[n^{1-k},1]$ does not increase the moment error. We allocate the individual failure probabilities so that one complete run succeeds with probability at least $2/3$.

Finally, $R\ge n^{1-k}$ implies
\[
 B\le O_k(n^{u_k}\eta^{-(1-u_k)}),\qquad
 R^{-1/(2k)}/\eta\le n^{b_k}/\eta.
\]
Substituting $\eta=\Theta_k(\eps)$ gives Eq.~\eqref{eq:main} at constant success probability. The initial search is absorbed by the first term. Repeating the complete algorithm $O(\log(1/\delta))$ times and taking the median gives success probability at least $1-\delta$. The overheads from circuit conversion and confidence amplification are polylogarithmic and are included in the stated bound.
\end{proof}

\bigskip

\appendix
\renewcommand{\thesection}{Appendix~\Alph{section}}

\section{Sparse collision tables for scale estimation}\label{app:sparse}
In this appendix, we prove Proposition~\ref{prop:sparse}, which is used in the binary search of Section~\ref{sec:scale}. Combining the proposition with that search and Eq.~\eqref{eq:log} also gives a direct entropy estimator using $\Oh_k(n^{u_k}/\eps)$ queries.

\subsection{Collision search with bounded populations}
For a table $x\in[n]^L$, write $N_y(x)$ for the multiplicity of label $y$ and define
\begin{equation}\label{eq:counts}
 B_i(x)=|\{y:N_y(x)\ge i\}|,\qquad
 C_i(x)=\sum_y\binom{N_y(x)}i.
\end{equation}
The quantities $B_i$ control the search cost, while $C_k$ is the statistic whose expectation we estimate.

\begin{lemma}[A capped-population search algorithm]\label{lem:search}
Fix $k\ge2$. Suppose a length-$L$ table has maximum multiplicity at most $K\ge k$, and known numbers satisfy
\[
 1\le D_{k-1}\le\cdots\le D_2\le L,
 \qquad B_i(x)\le D_i\quad(2\le i<k).
\]
The sequences are empty when $k=2$. Define
\begin{equation}\label{eq:Qsearch}
 \mathcal Q(L,D)=\max_{2\le j\le k}
 \left(L^{2^{j-1}}\prod_{i=2}^{j-1}D_i^{2^{i-2}}\right)^{1/(2^j-1)}.
\end{equation}
There is an algorithm that decides whether a $k$-collision exists and, when it exists, finds a collision label, with failure probability at most $\zeta$, using
\begin{equation}\label{eq:searchcost}
 \Oh_k(K^{(k-1)/2}\mathcal Q(L,D))
\end{equation}
queries. Its query count is fixed and its output is bounded even outside the promise. The suppressed factors are logarithmic in $L,n,1/\zeta$.
\end{lemma}

\begin{proof}
We use the colored dual-adversary construction in~\cite[Lemma 8.5 and Proposition 8.6]{PHD}, following its learning-graph approach. The positive inputs have a colorful $k$-collision and multiplicity at most $K$; negative inputs have no $k$-collision anywhere. Both domains obey the population promises. Thus each negative input has at most $k-1$ occurrences of any label.

Partition the positions randomly into $k$ balanced roles of sizes $v_i=\Theta_k(L)$. The number of equal-label prefixes across the first $j$ roles is at most $K^jD_j$. For raw set sizes $a_i$, the diagonal objective of the construction is, up to fixed constants,
\begin{equation}\label{eq:dualobjective}
 a_1+\sum_{j=2}^{k-1}a_j
 \sqrt{\frac{C_j^{\rm col}\prod_{i<j}a_i}{\prod_{i\le j}v_i}}
 +\sum_{s=1}^k\sqrt{\frac{\prod_{i\le s}v_i}{\prod_{i<s}a_i}},
 \qquad C_j^{\rm col}\le K^jD_j.
\end{equation}
We make explicit its dependence on $K$. Correction branches in the feasibility constraints count occurrences of the selected positive certificate label in a negative input. There are at most $k-1$, so their branch parameter is $k$ and the number of raw copies at level $i$ can be taken as $\ell_i=k^{i-1}(2^k-1)$. It depends only on $k$.

If the selected label has $d_i$ occurrences in role $i$, then $\sum_i d_i\le K$ and the positive normalization is
\[
 \gamma_x=\prod_{i<k}\frac{(v_i)_{d_i}}{(v_i-\ell_i a_i)_{d_i}},
\]
where $(z)_d$ denotes the falling factorial. Under $v_i\ge4K$ and $\ell_i a_i\le v_i/(8K)$,
\[
 \log\gamma_x\le2\sum_{i<k}\frac{d_i\ell_i a_i}{v_i}\le\frac14.
\]
Thus the correction multiplicities and positive normalization remain bounded by constants depending only on $k$. The cited feasibility construction retains the objective~\eqref{eq:dualobjective} with such constants. This establishes the required dependence on the multiplicity cutoff $K$.

To optimize, put $Q=\mathcal Q(L,D)$, $z_1=Q/L$, $a'_1=Q$, and, for $2\le j<k$, set
\begin{equation}\label{eq:searchparams}
 z_j=\min\left\{z_{j-1},\frac{Q\sqrt{z_{j-1}}}{\sqrt{LD_j}}\right\},
 \qquad a'_j=\frac{Lz_j}{z_{j-1}}.
\end{equation}
Since $D_j\le L$ and $L^{2/3}\le Q\le L$, these satisfy $Q\le a'_j\le L$ and
\[
 a'_j\sqrt{\frac{D_j\prod_{i<j}a'_i}{L^j}}\le Q.
\]
Expanding the minima shows that $z_{k-1}$ is the minimum, over subsequences $2\le i_1<\cdots<i_m\le k-1$ including the empty subsequence, of
\begin{equation}\label{eq:searchpaths}
 \frac{Q^{2-2^{-m}}}
 {L\prod_{t=1}^mD_{i_t}^{2^{-(m-t+1)}}}.
\end{equation}
For fixed $m$, its denominator is largest at $i_t=t+1$, since $D_i$ is nonincreasing. The $j=m+2$ term in Eq.~\eqref{eq:Qsearch} then implies that each expression is at least $L/Q^2$. Since $\prod_{i<s}a'_i=L^{s-1}z_{s-1}$, all certificate terms obey
\[
 \sqrt{L^s/\prod_{i<s}a'_i}\le Q.
\]
For $s=1$, this follows directly from $Q\ge L^{2/3}$.

Choose a sufficiently large fixed $c_k$ and take $a_i=\max\{1,\lfloor a'_i/(c_k{}K)\rfloor\}$. If $L\ge c'_k{}K$, the volume conditions hold, and $a'_i/(2c_k{}K)\le a_i\le a'_i$. Otherwise reading the table costs $O_k(K)$ and suffices. The setup terms in Eq.~\eqref{eq:dualobjective}, including $K^j$, and the enlarged certificate terms are all bounded by $O_k(K^{(k-1)/2}Q)$. The algorithmic characterization by the dual adversary bound~\cite{LMRSS} then gives the colored decision algorithm.

A fixed $k$-tuple is colorful with probability bounded away from zero in terms of $k$. We repeat the procedure with independent colorings and accept if any repetition finds a collision. This decides the existence of a collision with bounded error. A globally negative table is a valid negative input for every coloring; on positive tables a favorable coloring supplies a valid positive input. Repetition reduces the error to the required confidence.

For search, binary-search the label alphabet. At each decision step replace labels outside the selected label prefix by distinct position-tagged symbols. This uses $O(1)$ original lookups, preserves all population and multiplicity caps, and keeps a collision precisely when the selected prefix contains a collision label. After at most $\lceil\log_2n\rceil$ correct decisions, the remaining prefix identifies a collision label. We pad all operations to fixed query budgets, including executions on inputs outside the promise.
\end{proof}

\subsection{Moments and tails of a random table}
Let $X_1,\ldots,X_L$ be independent samples, with $L=\lceil k g^{-1/k}\rceil$ and $P_k(p)\le g\le1$. We use $B_i$ and $C_i$ for the corresponding random population and collision counts.

\begin{lemma}[Collision moments]\label{lem:collisionmoments}
There is a constant $A_k$ such that, for $2\le i\le k$ and every integer $h\ge1$,
\begin{equation}\label{eq:collisionmoments}
 \E C_i^h\le(A_k h)^{ih}\max\{1,L^iP_i(p)\}^h.
\end{equation}
Moreover,
\begin{equation}\label{eq:occupancy}
 L^iP_i(p)\le(k+1)^in^{1-i/k},\qquad
 \Pr\{\max_yN_y>K\}\le\frac{(k+1)^{K+1}}{(K+1)!}\quad(K\ge k-1).
\end{equation}
\end{lemma}
\begin{proof}
We have $Lp_{\max}\le k+1$, and Eq.~\eqref{eq:dimensionmoment} proves the first inequality in Eq.~\eqref{eq:occupancy}. Expand $C_i^h$ as a sum of $h$ collision constraints on $i$ distinct positions each. A connected component of overlapping constraints on $v\ge i$ positions forces all its labels to agree and contributes at most
\[
 L^vP_v(p)\le L^iP_i(p)(Lp_{\max})^{v-i}
 \le L^iP_i(p)(k+1)^{v-i}
\]
after summing its embeddings. There are at most $h$ components, at most $ih$ positions, and at most $(ih)^{ih}$ identification patterns. These bounds give Eq.~\eqref{eq:collisionmoments} after enlarging $A_k$.

If an occupancy exceeds $K$, then $C_{K+1}\ge1$. Thus
\[
 \Pr\{\max_yN_y>K\}
 \le\binom L{K+1}P_{K+1}(p)
 \le\frac{L^k{}P_k(p)(Lp_{\max})^{K+1-k}}{(K+1)!}
 \le\frac{(k+1)^{K+1}}{(K+1)!}.
\]
\end{proof}

\begin{corollary}[A promise with negligible truncation error]\label{cor:caps}
For $0<\eta\le1/4$, choose $h=\lceil c_k\log((n+2)/\eta)\rceil$ with sufficiently large $c_k$. There are caps
\begin{equation}\label{eq:caps}
 K=O_k(h),\qquad J=O_k(h^k),\qquad
 D_i=\min\{L,\lceil A n^{1-i/k}\rceil\},\quad A=O_k(h^k),
\end{equation}
with $D_2\ge\cdots\ge D_{k-1}\ge1$, such that
\begin{equation}\label{eq:badprob}
 \Pr\{\max_yN_y>K\text{ or some }B_i>D_i\}\le\frac\eta{100J},
 \qquad \E(C_k-J)_+\le\frac\eta{100}.
\end{equation}
The same conclusion holds with any larger fixed constants, depending only on $k$, in the denominators.
\end{corollary}
\begin{proof}
Apply Markov's inequality to Eq.~\eqref{eq:collisionmoments} with moment order $h$. A sufficiently large multiple of $h^k n^{1-i/k}$ gives tail probability at most $e^{-h}$. Taking the minimum with $L$ is valid because $B_i\le L$. The factorial estimate controls the maximum occupancy. Finally,
\[
 \E(C_k-J)_+\le\frac{\E C_k^h}{J^{h-1}}.
\]
Take $J$ to be a sufficiently large multiple of $(A_k h)^k$. The right-hand side is at most $Je^{-h}$ up to a fixed constant. Increasing $c_k$ absorbs the caps and the finite number of tail events.
\end{proof}

\subsection{The bounded collision-count estimator}
\begin{proof}[Proof of Proposition~\ref{prop:sparse}]
Set $L=\lceil k g^{-1/k}\rceil$ and $\beta=\binom{L}{k}$. Then
\begin{equation}\label{eq:meanC}
 \beta g=\Theta_k(1),\qquad \E C_k=\beta P_k(p)\le\beta g=O_k(1).
\end{equation}
Choose the caps in Corollary~\ref{cor:caps} with a sufficiently small fixed multiple of $\eta$. The following procedure computes a capped count on promised tables:
\begin{enumerate}
\item Initialize $c=0$ and an empty set of excluded labels.
\item Use Lemma~\ref{lem:search}, masking excluded labels by distinct position tags. If no collision remains, output $c$.
\item For the returned label, use quantum search~\cite{BBHT} to enumerate its positions and verify each reported position, stopping after $K+1$ discoveries or when a fixed query budget is reached. Under the promise there are at most $K$ positions; high-confidence search tests whether any remain.
\item If $d$ positions are found, add $\binom{d}{k}$ to $c$, cap the result at $J$, and exclude the label. Stop at the cap or after at most $J$ iterations; otherwise return to step 2.
\end{enumerate}
Every returned index is checked against the table. If a subroutine fails, a cap is exceeded, or the query budget is exhausted, return an arbitrary value in $[0,J]$. Reduce all decision and enumeration errors so their total probability is at most $\eta/(100J)$. Occurrence enumeration costs $\Oh(K\sqrt L)$ per iteration, which is absorbed, up to powers of $K$, by $\mathcal Q(L,D)\ge L^{2/3}$.

Let $Y\in[0,J]$ be the output. Combining its bounded range, the bad-table probability, and the clipping error gives
\begin{equation}\label{eq:Ybias}
 |\E Y-\beta P_k(p)|\le c'_k\eta,\qquad \E Y=O_k(1),
\end{equation}
where $c'_k$ can be made arbitrarily small by decreasing the fixed error tolerances.

We implement this procedure coherently by retaining its random bits, measurement records, and private table registers. Lemma~\ref{lem:table} gives the exact independent-table mean. Rotate a flag with conditional success probability $Y/J$. Its mean is $O_k(1/J)$, so amplitude estimation~\cite{BHMT} attains additive error $\Theta_k(\eta/J)$ using $O_k(\sqrt J/\eta)$ executions and inverse executions of the coherent procedure. Multiply the estimate by $J/\beta$. Eqs.~\eqref{eq:meanC}--\eqref{eq:Ybias} give error at most $\eta g$. The initial reflection includes the zero-initialized implicit table and is a known query-free operation.

For the execution cost, $g\ge P_k(p)\ge n^{1-k}$ gives $L\le(k+1)n^{1-1/k}$, and $D_i\le\Oh_k(n^{1-i/k})$. The exponent of $n$ in the $j$th term of~\eqref{eq:Qsearch} is
\begin{equation}\label{eq:dimensionexponent}
 \frac{2^{j-1}(1-1/k)+\sum_{i=2}^{j-1}2^{i-2}(1-i/k)}{2^j-1}
 =\frac{(3k-j)2^{j-2}-k}{k(2^j-1)}=u_{k,j}.
\end{equation}
All polynomial factors in $K,J,A$ and all confidence factors are polylogarithmic in $n/\eta$. Thus each execution costs $\Oh_k(n^{u_k})$, proving the stated $\Oh_k(n^{u_k}/\eta)$ bound. Median amplification gives any desired confidence. The explicit caps bound both the output and the query cost on every execution.
\end{proof}

\section{A simpler order-two estimator}\label{app:two}
Algorithm~\ref{alg:main} applies to $k=2$, with no intermediate collision-state levels. In this section, we give a simpler implementation for this order. It uses only the first mean-estimation step of Lemma~\ref{lem:prefix}, both to locate the power sum and to obtain the final estimate.

\begin{proposition}[Direct order-two estimator]\label{prop:two}
Let $P=P_2(p)$ and suppose $P\le R\le CP$, $R\le1$, for fixed $C$. For sufficiently small $\eps>0$, the order-one to order-two prefix mean primitive estimates $H_2(p)$ to additive error $\eps$ using
\begin{equation}\label{eq:twoinstance}
 \Oh\!\left(R^{-1/3}\eps^{-2/3}+R^{-1/4}/\eps\right)
\end{equation}
ordinary purified queries at constant success probability. The same primitive finds the initial scale at worst-case cost $\Oh(n^{1/3})$. The complete procedure is given in Algorithm~\ref{alg:two}. Hence
\begin{equation}\label{eq:two}
 Q_2(n,\eps)\le\Oh(n^{1/3}\eps^{-2/3}+n^{1/4}/\eps).
\end{equation}
\end{proposition}

\begin{algorithmblock}{Quantum algorithm for order-two \Renyi{} entropy estimation}\label{alg:two}
\textbf{Input:} Purified access oracle $U_p$; dimension bound $n$; additive accuracy $\eps$; maximum failure probability $\delta\in(0,1/3)$.\par
\textbf{Output:} An estimate $\widetilde H_2(p)$ with additive error at most $\eps$ and failure probability at most $\delta$.
\begin{enumerate}
\item Start with $g=1$. At each candidate, set $\varphi=\sqrt g$ and $L=\lceil g^{-1/3}\rceil$. Use the $s=1$ mean primitive of Lemma~\ref{lem:prefix} to estimate $P_2(p)$ to additive error $g/16$.
\item If the estimate is at least $g/4$, retain $R=g$. Otherwise replace $g$ by $\max\{g/2,1/n\}$ and repeat, subject to a fixed limit of $O(\log n)$ tests.
\item Set $\varphi=\sqrt R$ and $L=\lceil R^{-1/3}\eps^{-2/3}\rceil$. Prepare $L$ raw samples and use the same primitive to estimate $P_2(p)$ to additive error $c\eps R$, for a sufficiently small constant $c$.
\item Project the result onto $[1/n,1]$ and take its negative logarithm.
\item Repeat the complete procedure $O(\log(1/\delta))$ times and return the median.
\end{enumerate}
The confidence assignments in steps 1--4 make each complete run succeed with probability at least $2/3$.
\end{algorithmblock}

\begin{proof}
Use the whole alphabet as both sets in Lemma~\ref{lem:prefix}, with
\[
 s=1,\quad M_1=1,\quad D_1=O(1),\quad G=P,\quad
 g=R,\quad\varphi=\sqrt{R}.
\]
The promises hold because $p_{\max}^2\le P\le R$ and $R\le\sqrt{R}=M_1\varphi$. Prepare a cache of $L$ raw samples and the constant number of additional source samples. Comparing the selected sample with a fresh sample gives a match probability of $P$, and the source normalization is exactly one.

With additive moment error $\Delta=c'\eps R$, Eq.~\eqref{eq:uppermean}, including setup, gives
\begin{align}
 Q(L)&\le\Oh\!\left(L+\frac{\sqrt{\varphi R}+\sqrt{R/L}}{\eps R}\right)\notag\\
 &=\Oh\!\left(L+\frac{R^{-1/4}}\eps+
                   \frac1{\eps\sqrt{LR}}\right).\label{eq:twobalance}
\end{align}
Choose $L=\lceil R^{-1/3}\eps^{-2/3}\rceil$. The first and last terms are then $O(R^{-1/3}\eps^{-2/3})$, proving Eq.~\eqref{eq:twoinstance}. Compose the cache initialization and mean transducer before circuit conversion. Projection onto $[1/n,1]$ and the negative logarithm give the entropy estimate.

To obtain the initial scale, start at $g=1$. At each valid candidate upper bound, set $\varphi=\sqrt g$, $L=\lceil g^{-1/3}\rceil$, and estimate $P$ to error $g/16$ by the same primitive. Its cost is $\Oh(g^{-1/3}+g^{-1/4})=\Oh(g^{-1/3})$. Stop if the estimate is at least $g/4$; otherwise replace $g$ by $\max\{g/2,1/n\}$. When the estimate is accurate, continuation preserves the upper-bound assumption and stopping gives $P\ge3g/16$. Since $P\ge1/n$, an accurate test must stop at $g=1/n$.

Fix the number of tests and their caps using $P\ge1/n$, and amplify their confidence according to that number. The geometric cost sum is at most $\Oh(n^{1/3})$. If the intervals become inconsistent or the query budget is exhausted, we declare failure. The returned scale obeys $P\le R\le(16/3)P$. Substituting $R\ge1/n$ in Eq.~\eqref{eq:twoinstance} proves~\eqref{eq:two}.
\end{proof}

Thus the same prefix framework applies at every fixed integer order. Order two uses raw samples as its only cache and has no intermediate moment normalizations. It does not require label collection or the batch estimator.


\medskip
\begin{thebibliography}{99}
\raggedright

\bibitem{Renyi}
A.~R\'enyi, ``On measures of entropy and information,'' in \emph{Proc. 4th Berkeley Symp. Mathematical Statistics and Probability}, vol.~1, pp.~547--561, 1961.

\bibitem{AOST}
J.~Acharya, A.~Orlitsky, A.~T.~Suresh, and H.~Tyagi, ``Estimating R\'enyi entropy of discrete distributions,'' \emph{IEEE Trans. Inf. Theory}, vol.~63, no.~1, pp.~38--56, 2017. \href{https://arxiv.org/abs/1408.1000}{arXiv:1408.1000}.

\bibitem{LiWu}
T.~Li and X.~Wu, ``Quantum query complexity of entropy estimation,'' \emph{IEEE Trans. Inf. Theory}, vol.~65, no.~5, pp.~2899--2921, 2019. \href{https://doi.org/10.1109/TIT.2018.2883306}{doi:10.1109/TIT.2018.2883306}.

\bibitem{GL}
A.~Gily\'en and T.~Li, ``Distributional property testing in a quantum world,'' in \emph{Proc. 11th Innovations in Theoretical Computer Science Conf. (ITCS)}, LIPIcs, vol.~151, pp.~25:1--25:19, 2020. \href{https://doi.org/10.4230/LIPIcs.ITCS.2020.25}{doi:10.4230/LIPIcs.ITCS.2020.25}.

\bibitem{BJY}
A.~Belovs, S.~Jeffery, and D.~Yolcu, ``Taming quantum time complexity,'' \emph{Quantum}, vol.~8, Art.~no.~1444, 2024. \href{https://doi.org/10.22331/q-2024-08-23-1444}{doi:10.22331/q-2024-08-23-1444}.

\bibitem{CAG}
A.~Cornelissen, S.~Apers, and S.~Gribling, ``Improved quantum volume estimation with transducers and amortized quantum walks,'' 2026. \href{https://arxiv.org/abs/2609.40217v2}{arXiv:2609.40217v2}.

\bibitem{Multi}
K.~Chen, M.~Gao, T.~Li, Q.~Wang, and X.~Wang, ``Quantum multi-level estimation of functionals of discrete distributions,'' in \emph{Proc. 53rd Int. Colloq. Automata, Languages, and Programming (ICALP)}, LIPIcs, vol.~374, pp.~58:1--58:23, 2026. \href{https://doi.org/10.4230/LIPIcs.ICALP.2026.58}{doi:10.4230/LIPIcs.ICALP.2026.58}. Full version: \href{https://arxiv.org/abs/2605.03685v1}{arXiv:2605.03685v1}.

\bibitem{PHD}
T.~Peng, S.~Hou, and X.~Dong, ``Tight quantum query bounds for collision finding over arbitrary bounded types,'' 2026. \href{https://arxiv.org/abs/2611.17804v1}{arXiv:2611.17804v1}.

\bibitem{BHMT}
G.~Brassard, P.~H\o yer, M.~Mosca, and A.~Tapp, ``Quantum amplitude amplification and estimation,'' in \emph{Quantum Computation and Quantum Information}, Contemporary Mathematics, vol.~305, pp.~53--74, American Mathematical Society, 2002. \href{https://arxiv.org/abs/quant-ph/0005055}{arXiv:quant-ph/0005055}.

\bibitem{Montanaro}
A.~Montanaro, ``Quantum speedup of Monte Carlo methods,'' \emph{Proc. R. Soc. A}, vol.~471, no.~2181, Art.~no.~20150301, 2015. \href{https://arxiv.org/abs/1504.06987}{arXiv:1504.06987}.

\bibitem{vA}
J.~van Apeldoorn, ``Quantum probability oracles \& multidimensional amplitude estimation,'' in \emph{Proc. 16th Conf. Theory of Quantum Computation, Communication and Cryptography (TQC)}, LIPIcs, vol.~197, pp.~9:1--9:11, 2021. \href{https://doi.org/10.4230/LIPIcs.TQC.2021.9}{doi:10.4230/LIPIcs.TQC.2021.9}.

\bibitem{CWZ}
K.~Chen, Q.~Wang, and Z.~Zhang, ``A list of complexity bounds for property testing by quantum sample-to-query lifting,'' 2025. \href{https://arxiv.org/abs/2512.01971v1}{arXiv:2512.01971v1}.

\bibitem{Belovs}
A.~Belovs, ``Learning-graph-based quantum algorithm for $k$-distinctness,'' in \emph{Proc. 53rd Annu. IEEE Symp. Foundations of Computer Science (FOCS)}, pp.~207--216, 2012.

\bibitem{LMRSS}
T.~Lee, R.~Mittal, B.~W.~Reichardt, R.~\v{S}palek, and M.~Szegedy, ``Quantum query complexity of state conversion,'' in \emph{Proc. 52nd Annu. IEEE Symp. Foundations of Computer Science (FOCS)}, pp.~344--353, 2011. \href{https://arxiv.org/abs/1011.3020}{arXiv:1011.3020}.

\bibitem{YLC}
T.~J.~Yoder, G.~H.~Low, and I.~L.~Chuang, ``Fixed-point quantum search with an optimal number of queries,'' \emph{Phys. Rev. Lett.}, vol.~113, Art.~no.~210501, 2014. \href{https://arxiv.org/abs/1409.3305}{arXiv:1409.3305}.

\bibitem{BBHT}
M.~Boyer, G.~Brassard, P.~H\o yer, and A.~Tapp, ``Tight bounds on quantum searching,'' \emph{Fortschr. Phys.}, vol.~46, no.~4--5, pp.~493--505, 1998. \href{https://arxiv.org/abs/quant-ph/9605034}{arXiv:quant-ph/9605034}.

\end{thebibliography}
\end{document}